\documentclass[aps,reprint,superscriptaddress,longbibliography,prb]{revtex4-1}

\usepackage{graphicx, amsmath, verbatim, dsfont, amsfonts, color}
\usepackage{braket}
\usepackage[us]{datetime}

\begin{document}

\title{Parafermion supporting platform based on spin transitions in the fractional quantum Hall effect regime}

\author{Tailung Wu}
\thanks{These authors contributed equally to this work}
\affiliation{Department of Physics and Astronomy, Purdue University, West Lafayette, IN 47907 USA}
\affiliation{Birck Nanotechnology Center, Purdue University, West Lafayette, IN 47907 USA}
\author{Aleksandr Kazakov}
\thanks{These authors contributed equally to this work}
\affiliation{Department of Physics and Astronomy, Purdue University, West Lafayette, IN 47907 USA}
\author{George Simion}
\affiliation{Department of Physics and Astronomy, Purdue University, West Lafayette, IN 47907 USA}
\author{Zhong Wan}
\affiliation{Department of Physics and Astronomy, Purdue University, West Lafayette, IN 47907 USA}
\author{Jingcheng Liang}
\affiliation{Department of Physics and Astronomy, Purdue University, West Lafayette, IN 47907 USA}
\author{Kenneth W. West}
\affiliation{Department of Electrical Engineering, Princeton University, Princeton, NJ 08540 USA}
\author{Kirk Baldwin}
\affiliation{Department of Electrical Engineering, Princeton University, Princeton, NJ 08540 USA}
\author{Loren N. Pfeiffer}
\affiliation{Department of Electrical Engineering, Princeton University, Princeton, NJ 08540 USA}
\author{Yuli Lyanda-Geller}
\affiliation{Department of Physics and Astronomy, Purdue University, West Lafayette, IN 47907 USA}
\affiliation{Birck Nanotechnology Center, Purdue University, West Lafayette, IN 47907 USA}
\author{Leonid P. Rokhinson}
\affiliation{Department of Physics and Astronomy, Purdue University, West Lafayette, IN 47907 USA}
\affiliation{Birck Nanotechnology Center, Purdue University, West Lafayette, IN 47907 USA}
\affiliation{Department of Electrical and Computer Engineering, Purdue University, West Lafayette, IN 47907 USA}

\date{September 22, 2017}

\begin{abstract}
We propose an experimentally-feasible system based on spin transitions in the fractional quantum Hall effect regime where parafermions, high-order non-abelian excitations, can be potentially realized. We provide a proof-of-concept experiments showing that in specially designed heterostructures spin transitions at a filling factor 2/3 can be induced electrostatically, allowing local control of polarization and on-demand formation of helical domain walls with fractionalized charge excitations, a pre-requisite ingredient for parafermions formation. We also present exact diagonalization numerical studies of domain walls formed between domains with different spin polarization in the fractional quantum Hall effect regime and show that they indeed possess electronic and magnetic structure needed for parafermion formation when coupled to an s-wave superconductor.
\end{abstract}

\maketitle

Topological quantum computation can be performed with Majorana fermions (MF)\cite{Nayak2008}, but MF-based qubits are not computationally universal\cite{Baraban2010}. Parafermions (PFs), higher order non-Abelian excitations, are predicted to have denser rotation group and their braiding enables two-qubit entangling gates\cite{Fendley2012,Alicea2016}. A two-dimensional array of parafermions can serve as a building block for the system which supports Fibonacci anyons with universal braiding statistics\cite{Mong2013}, a holy grail of topological quantum computing.
In an important conceptual paper Clark {\it et al.} proposed that PF excitations can emerge in the fractional quantum Hall effect (FQHE) regime if two counter-propagating fractional chiral edge states with opposite polarization are brought into close proximity in the presence of superconducting coupling\cite{Clarke2012}. Here we propose that domain walls formed during spin phase transitions in the FQHE regime have the prerequisite helical structure to support PF excitations when coupled to an s-wave superconductor. We demonstrate experimentally that in a triangular quantum well a 2D system can be tuned across a spin transition at a filling factor $\nu=2/3$ using electrostatic gating. Such local control of polarization allows formation of a reconfigurable network of helical domain walls with fractionalized charge excitations (hDW) along gate boundaries and, potentially, parafermion manipulation and braiding. We present exact  diagonalization numerical studies of domain walls formed between domains with different polarization and calculate their electronic and magnetic structure.

Helical channels are commonly associated with the quantum spin Hall effect\cite{Maciejko2011}, topological insulators\cite{Hasan2011} or nanowires with spin-orbit interactions\cite{Lutchyn2010a,Oreg2010}, where Coulomb interactions are not strong enough to fractionalize charges. A natural system to look for PFs is a 2D electron gas (2DEG) in the FQHE regime, where  edge states support fractionally charged excitations. In the conventional QHE setting, though, the edge modes are \textit{chiral}. \textit{Helical} channels can potentially emerge as domain walls during a quantum Hall ferromagnetic transition.
It has been predicted that domain walls formed in the integer QHE regime at a filling factor $\nu=1$ have helical magnetic order\cite{falko00}. Experimentally, local electrostatic control of domain walls in the integer QHE regime at $\nu=2$ was recently demonstrated in magnetic semiconductors\cite{Kazakov2017a}, and their electronic and magnetic structure has been calculated \cite{Simion2017}.

In the FQHE regime spin transitions have been observed at a filling factor $\nu=2/3$ and other fractions\cite{Eisenstein1990,Smet2001}. At the transition, the 2DEG spontaneously phase separates into regions of different spin polarizations, and conducting domain walls are formed along the domain boundaries\cite{Verdene2007,Hayakawa2012}. An experimental challenge is to devise a system where spin transitions in the FQHE regime can be controlled locally, allowing formation and manipulation of DWs. Theoretically, neither magnetic nor electronic structure of these domain walls is known. Naively, a $\nu=2/3$ edge state can be viewed as two non-interacting co-propagating $\nu=1/3$ chiral edge modes, so that a domain wall can be constructed from two counter-propagating modes with opposite spin polarization, similar to the domain walls formation in the integer quantum Hall ferromagnetic transition\cite{Simion2017}. It has been realized, though, that unpolarized $\nu=2/3$ edge states are complex objects with chiral downstream propagation of charge excitations and neutral energy-carrying modes propagating both up- and downstream\cite{Kane1994,Bid2010,Venkatachalam2012}. Thus, co-propagating edge charge modes are strongly interacting and cannot be simply spatially separated to form a domain wall in the bulk. In the following we demonstrate a system where spin polarization of a $\nu=2/3$ FQHE state can be controlled by electrostatic gating. We also present a theory of electrostatically-induced domain walls formed between coexisting spin-polarized and unpolarised phases at $\nu=2/3$ and discuss their electronic and magnetic structure.

\begin{figure}
\centering\includegraphics[width=\columnwidth]{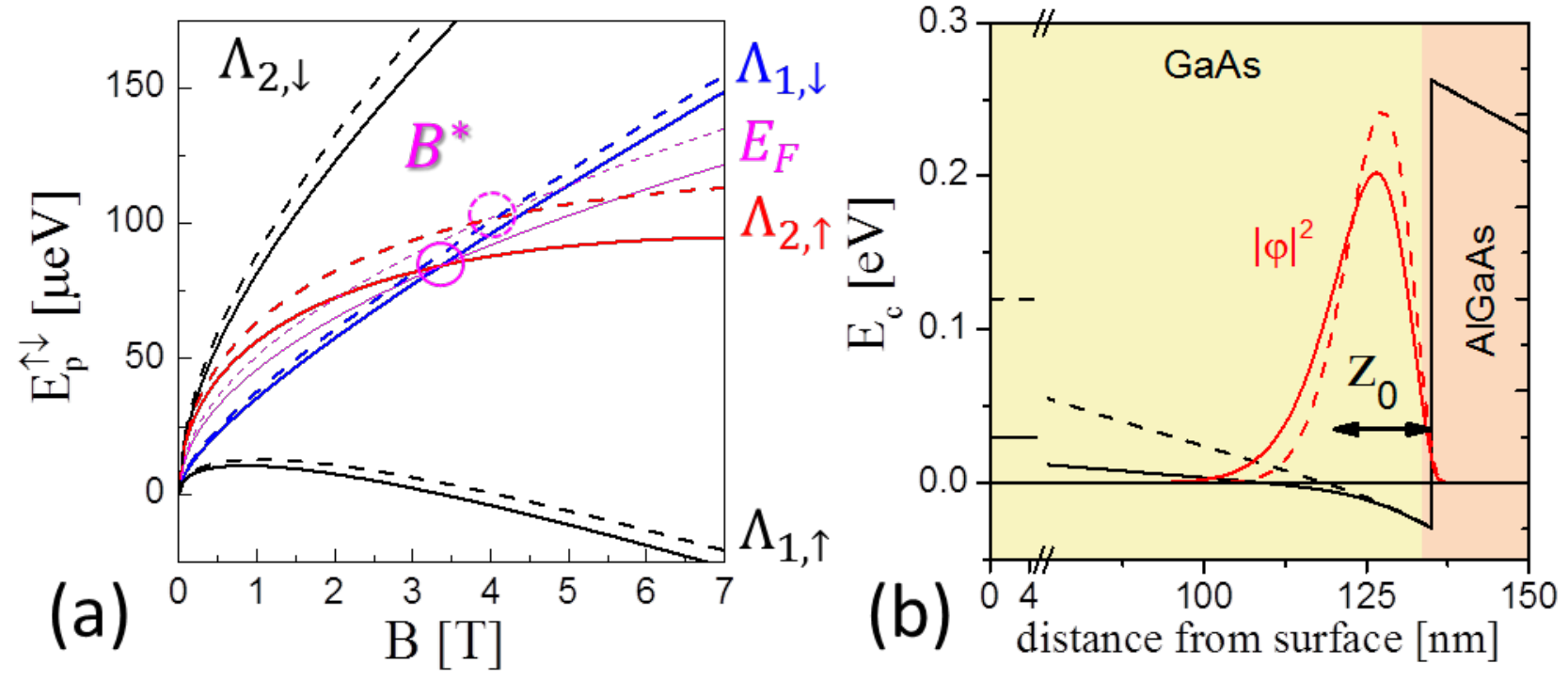}
\vspace{0in}
\caption{\textbf{Gate control of the wavefunction extent in a triangular well and energy spectrum of composite fermions.} (a) Energy spectrum of $\Lambda$-levels for CFs, Eq.~\ref{Ecf}. For $\nu=2/3$ (two filled $\Lambda$-levels) the spin polarization of the top level changes at $B^*$, when $\Lambda_{1,\downarrow}$ and $\Lambda_{2,\uparrow}$ cross. Solid and dotted lines are calculated for two different values of the wavefunction extent $z_0$. (b) The calculated wavefunction in a triangular quantum well formed at a GaAs/AlGaAs heterojunction interface. Solid and dotted lines correspond to the two different gate voltages and show the change of $z_0$. Note the break in the horizontal axis.}
\label{f:fig1}
\vspace{-0.2in}
\end{figure}

Spin transitions in the FQHE regime can be readily understood within the framework of the theory of composite fermions (CF)\cite{JainCFbook2007}, where FQHE states at filling factors $\nu=\nu^*/(2\nu^*-1)$ for $1/2<\nu<1$ are mapped onto integer QHE states with a filling factor $\nu^*$ for CFs. The energy spectrum of CF $\Lambda$-levels with an index $p=1, 2, 3 ...$ can be written as
\begin{equation}
E_p^{\uparrow\downarrow}=\hbar\omega_c^{cf}(p-\frac12)\pm g\mu_BB.
\label{Ecf}
\end{equation}
The CF cyclotron energy $\hbar\omega_c^{cf}$ is proportional to the charging energy $E_c=e^2/\sqrt{l_m^2+z_0^2}$, where $l_m\propto\sqrt{B_{\bot}}$ is the magnetic length, $B_{\bot}=B\cos\theta$ is the out-of-plane component of the magnetic field $B$, and $z_0$ is the extent of the wavefunction in the out-of-plane direction. The second term is the Zeeman energy. Due to the difference in $B$-dependences of the two terms, levels $\Lambda_{p,\downarrow}$ and $\Lambda_{p+1,\uparrow}$ cross at some $B^*>0$, see Fig.~\ref{f:fig1}. Thus, for $\nu^*\ge2$ (two or more $\Lambda$-levels are filled) the top energy level undergoes a spin transition at $B^*$.

Conventionally, FQHE spin transitions are studied in tilted magnetic fields, where \textit{global} control of the field angle $\theta$ changes the ratio of Zeeman and cyclotron energies. For a triangular confinement, though, $z_0$ is gate dependent, $z_0=z_0(V_g)$ (see Fig.~\ref{f:fig1}b), and \textit{local} control of $E_c$ and $B^*$ at a fixed $B$ becomes possible. Within the Fang-Howard approximation of the wavefunction in a triangular well $z_0=3/b$, where $b\propto n^{1/3}$ is a function of electron density. For GaAs parameters and $B^*\approx B_{\nu=2/3}\approx 4-6$ T, the field $B^*$ becomes density- and gate-dependent: $\delta B^*/B^*\approx 0.3 \delta n/n$, $\delta n/n=\delta V_g/V_g$. The field position of the $\nu=2/3$ state is also density- and gate-dependent, $\delta B_{\nu=2/3}/B_{\nu=2/3}=\delta n/n$. Thus, for a well-developed wide $\nu=2/3$ state and a sharp spin transition, there should be a range of magnetic fields where spin polarization of the top level can be tuned by a local electrostatic gating.

\begin{figure}[t]
\centering\includegraphics[width=\columnwidth]{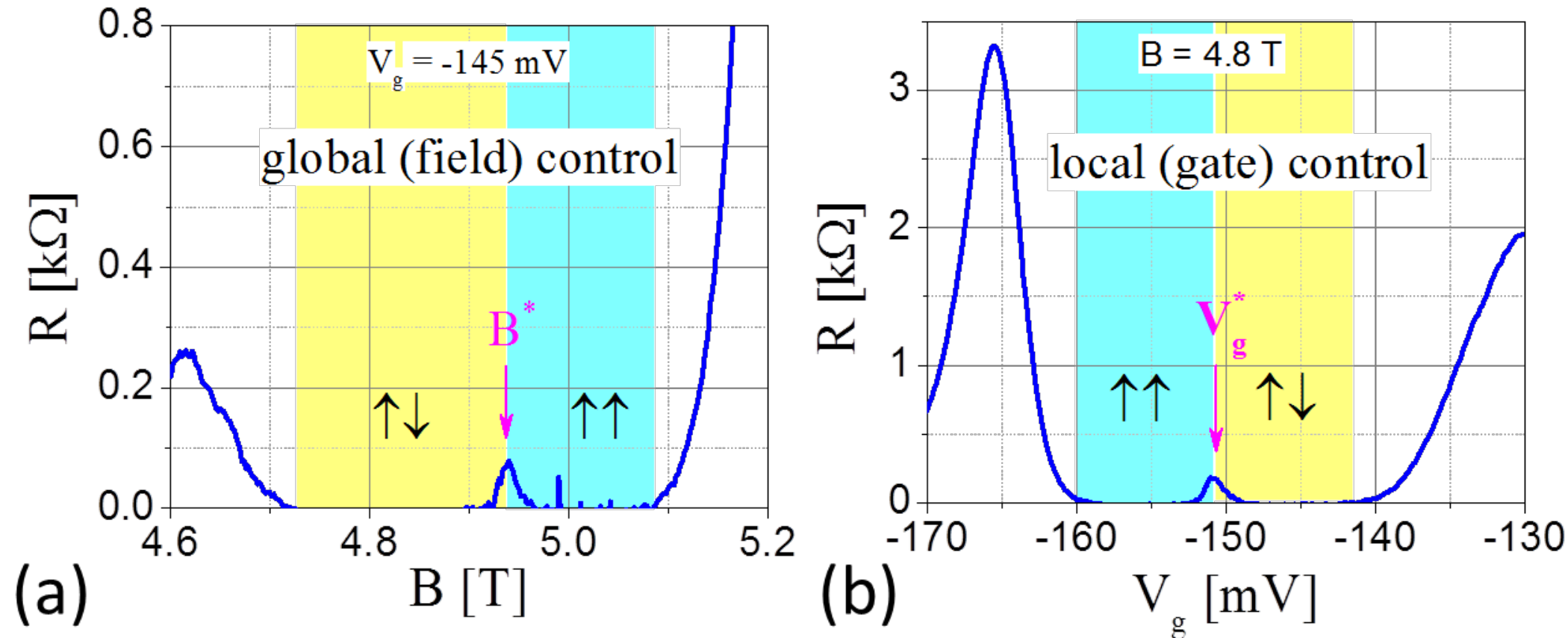}
\vspace{-0.2in}
\caption{\textbf{Field and gate control of FQHE spin transition.} Resistance near the $\nu=2/3$ state is measured as a function of (a) magnetic field $B$ at a constant gate voltage $V_g$ or (b) as a function of $V_g$ at a constant $B$ in a Hall bar sample from wafer A. In yellow regions the 2D gas is unpolarized with CFs levels $\Lambda_{1,\uparrow}$ and $\Lambda_{1,\downarrow}$ filled. In cyan regions 2D gas is fully polarized with CFs residing in $\Lambda_{1,\uparrow}$ and $\Lambda_{2,\uparrow}$ levels. $B^*$ and $V_g^*$ mark the spin transition.}
\label{f:fig2}
\vspace{-0.2in}
\end{figure}

In order to demonstrate electrostatic control of polarization we have grown a number of wafers where high mobility 2D electron gas is confined at a single GaAs/AlGaAs interface. Devices are patterned into Hall bars or quasi-Corbino geometry, and electron density is controlled by an electrostatic gate. Magnetoresistance in the vicinity of a $\nu=2/3$ plateau is shown in Fig.~\ref{f:fig2}a. At the base temperature of $\approx 30$ mK, the plateau is 0.35 T wide and is interrupted by a small peak at $B^*=4.94$ T. This peak has all the characteristics of a spin phase transition studied in the past, including strong current dependence and hysteresis with respect to the field sweep direction, which appears at high bias currents\cite{Smet2001}. We identify this peak with the spin transition in the topmost CFs energy level, when $\Lambda_{1,\downarrow}$ and $\Lambda_{2,\uparrow}$ levels cross. This is a transition between fully unpolarized and fully polarized $\nu=2/3$ states marked with yellow and cyan on the plot. The most important data is shown in Fig.~\ref{f:fig2}b, where resistance is plotted as a function of a gate voltage $V_g$ measured at a fixed magnetic field. Similarly to the field scan, a 18 mV - wide $\nu=2/3$ plateau is interrupted by a small peak at $V_g^*=-151$ mV, which we identify with the spin transition. The transition peak is narrow, $\approx 6$ mV with well-defined $\nu=2/3$ states with different polarization on the two flanks of the peak. Thus, it is possible to control spin polarization and induce spin transition in the topmost CF $\Lambda$-level by local electrostatic gating.

\begin{figure}[t]
\centering\includegraphics[width=\columnwidth]{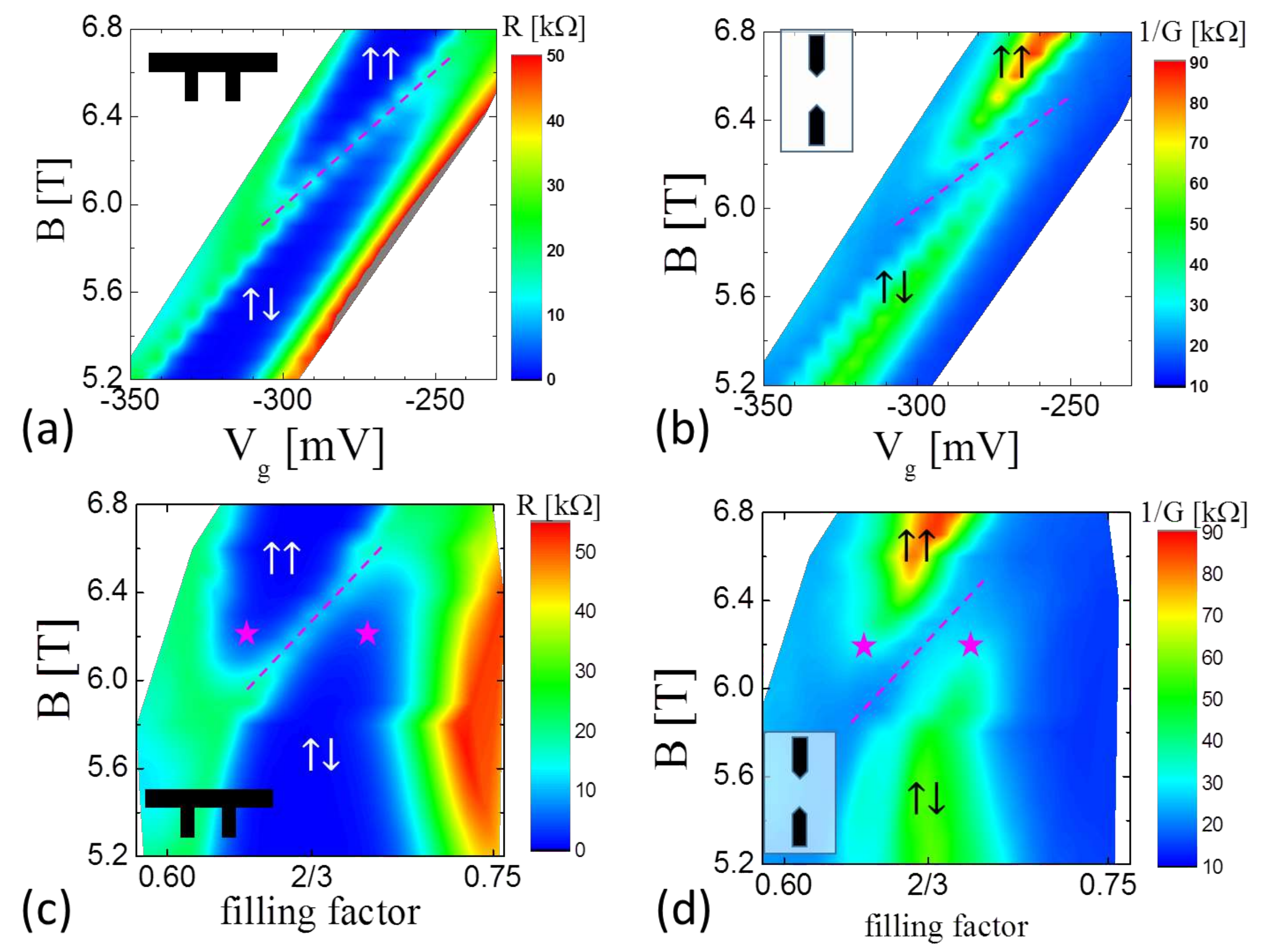}
\vspace{-0.2in}
\caption{\textbf{Evolution of the spin transition at $\nu=2/3$ in the field-gate voltage plane}. Spin transitions are measured (a) in a Hall bar and (b) in a quasi-Corbino geometry samples fabricated from wafer B. In (c,d) the same data as in (a,b) is re-plotted as a function of a filling factor. The spin transition is highlighted with a dashed line. In the 6.1~T~$<B<6.4$ T range both spin unpolarized and polarized states are accessible with the gate voltage, the optimum points with the lowest bulk conduction are marked with stars. Note that in the Corbino geometry inverse conductivity $1/G$ is measured; $G\rightarrow0$ at $\nu=2/3$ far from the spin transition.}
\label{f:fig3}
\vspace{-0.2in}
\end{figure}

An evolution of the spin transition in the $B-V_g$ and $B-\nu$ planes is shown in Fig.~\ref{f:fig3} for a Hall bar and quasi-Corbino geometry samples. In the Corbino sample ohmic contacts are 5 $\mu$m apart and the measured quantity is the inverse conductance $1/G$. At $\nu=2/3$ far from the transition $G\rightarrow0$ and $1/G$ is high. At a spin transition a random network of conducting domain walls is formed between the contacts and $1/G$ is reduced. There is a fundamental difference between the two geometries: in a Hall bar the edge corresponding to the lowest $\Lambda_{1,\uparrow}$ energy level always connects the contacts and non-zero resistance is the result of backscattering between edge channels along domain walls. In the Corbino geometry, contacts are not connected to the $\Lambda_{1,\uparrow}$ edge channel, and the measured current is carried by the domain walls formed from the counter-propagating  $\Lambda_{1,\downarrow}$ and $\Lambda_{2,\uparrow}$ states.

\begin{figure}
\includegraphics[width=0.98\columnwidth]{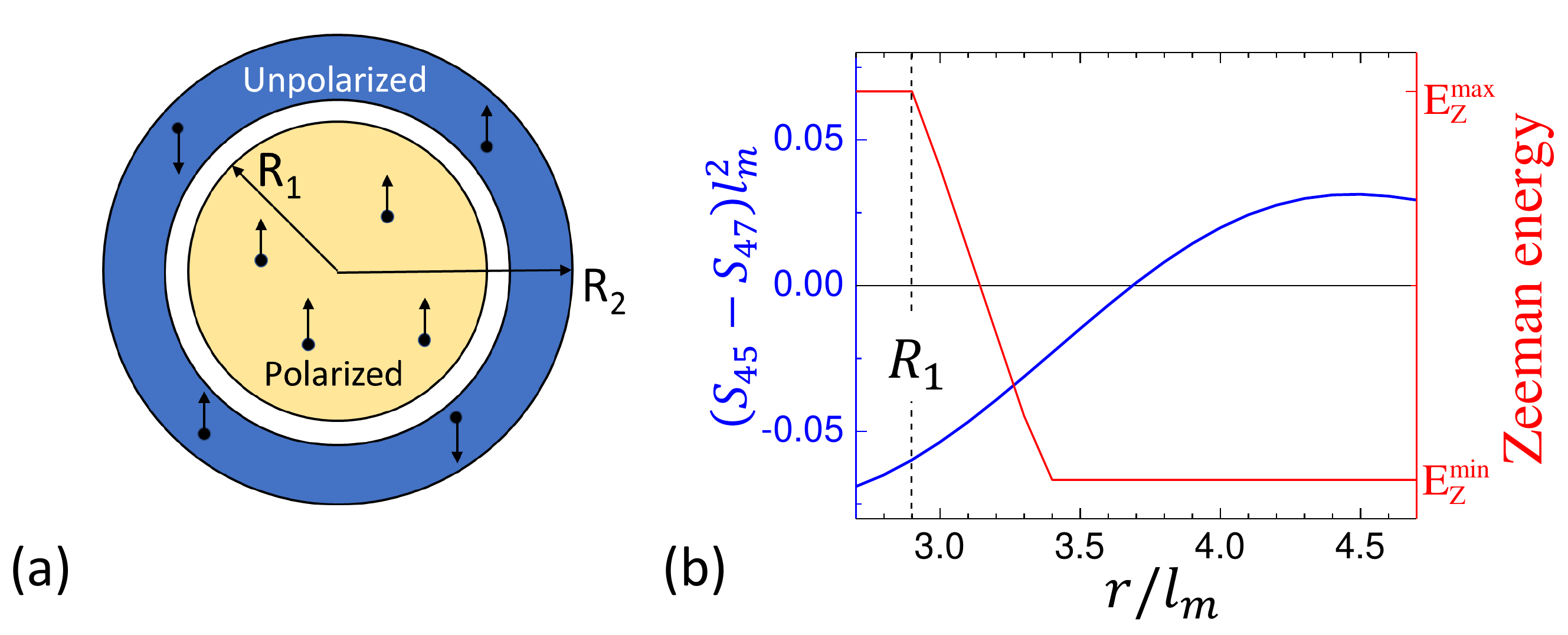}
\caption{\textbf{Simulation of edge states near electrostatically induced boundary.} (a) Disk geometry for the simulation domain. (b) The blue curve is the difference  between spin densities $S_z(r)$ for the modes with angular momentum $L_z=45$ and $47$, the current-caring exciting states on the two sides of the domain wall formed around $R_1$. Profile of the spatially-dependent Zeeman interaction used to form the domain wall is shown in red.}
\label{f:sim}
\vspace{-0.2in}
\end{figure}

In order to investigate the structure of domain walls formed between spin-unpolarized and spin-polarized regions we performed exact diagonalization studies of a system with small number of particles.
To simulate edge states we use the disk geometry\cite{Rezayidisk,TylanTyler2017} shown schematically in Fig.~\ref{f:sim}. Long-range Coulomb interactions between electrons are introduced using Haldane pseudopotentials. A neutralizing background and a confinement potential are used to hold electrons inside the disk. As is evident from experiments, spin transition can be controlled by modulation of either Coulomb or Zeeman energies interchangeably, see e.g. Fig.~\ref{f:fig3}. In our modeling we use spatially-dependent Zeeman effect to control spin polarization of the 2DEG. The central region of the disk of radius $R_1=2.9 l_m$ is characterized by a high Zeeman energy term $E_Z^{max}$, while the outer region with the outer diameter $R_2=4.8 l_m$ is set to $E_Z^{min}=0$. The Zeeman term varies smoothly within $R_1< r <R_1+\Delta R$, where $\Delta R=0.4 l_m$, resulting in a smooth variation of wavefunctions across the disk and avoiding spurious effects originating from abrupt changes. Note that due to a strong penetration of electron wavefunctions from the outer $R_1<r<R_2$ region into the inner $r<R_1$ region, the variation of the average spin splitting $\int\psi(r)^*E_Z(r)\psi(r)d^2r$ for electronic modes on opposite sides of the domain wall is $<6\%$, similar to the experimental conditions.  Therefore, this model approximate soft edges characterizing the bulk domain walls.

 We include up to 12 electrons in the exact diagonalization calculation for a fully spin-polarized states and 8 electrons for unpolarized states or coexisting polarized and unpolarized states at $\nu=2/3$. Energies and wavefunctions for the ground state and edge states, their electron and spin density distributions for the disk geometry are calculated, see Supplementary Materials for details.
The ground state for 8 particles has the total angular momentum projection $L_z=46$, in agreement with the composite fermion theory\cite{JainCFbook2007}.
The ground state is spin polarized in the interior part of the disk and spin-unpolarized in the exterior area, as expected. The total spin projection of the ground state is $S_z=+2$. The lowest excited states with the same spin projection, which correspond to the addition or subtraction of a single flux, have angular momenta $L_z=45$ and $47$. These states are the current carrying states defining the domain wall. The difference in spin polarizations between these two states in shown in Fig.~\ref{f:sim}b, it smoothly changes sign across the domain wall. There is $\approx 0.5 l_m$ outward shift of the position of the midpoint of the domain wall in the actual spin density profile relative to the profile of the defining Zeeman term. This shift is due to smaller wavefunction weights in the outer region compared to the inner region.

 The two edge states with different $L_z$ on the disk have different angular velocities. When mapped onto a plane, these two states will have different linear velocities, i.e. their velocities will have counter-propagating components. Combined with the different spin polarization these states will have a finite overlap with Cooper pair wavefunctions when coupled to an s-wave superconductor, as has been shown for domain walls in integer quantum Hall ferromagnets \cite{Simion2017}. Due to higher degeneracy of the composite fermion $\Lambda$-levels compared to the degeneracy of the Landau levels, parafermion states  will emerge at the boundary of topologically trivial s-wave  superconductor and a topologically non-trivial superconductor proximity-induced in the domain wall, as predicted in the Ref.~[\onlinecite{Clarke2012}].

\vspace{0.5cm}
\textbf{Methods}\\
Devices are fabricated from MBE-grown inverted GaAs/AlGaAs heterojunctions with 130-230 nm top GaAs layer and Si $\delta$-doping placed 70-300nm beneath the heterojunction interface. The top 25nm of GaAs are lightly doped to reduce the surface pinning potential. The 0.3 K electron mobilities are in the $\mu = 2-5\cdot10^{6}$ cm$^{2}$/Vs range for densities $n=0.2-1.6\cdot10^{11}$ cm$^{-2}$. The data reported in the main text is measured on wafers A and B with $n=1.3$ and $1.6\cdot10^{11}$ cm$^{-2}$ respectively, wafers parameters can be found in the Supplementary Material. Inverted heterostructures allow electrostatic gating of a shallow 2D gas with no hysteresis, also in similar wafers proximity-induced superconductivity has been reported\cite{Wan2015}. Devices are patterned into Hall bar or quasi-Corbino geometry, where contacts are placed inside the mesa and do not overlap mesa boundaries. Ohmic contacts are formed by annealing Ni/AuGe/Ni/Au 6nm/120nm/20nm/20nm in a H$_2$/N$_2$ atmosphere. Electrostatic gates are formed by evaporating 10 nm of Ti on a 50 nm Al$_2$O$_3$ isolation layer grown by an atomic layer deposition (ALD).
The gates are semitransparent and a 2D electron gas was created by shining red LED at $\sim 4$ K. Measurements were performed in a dilution refrigerator using standard 4-probe constant current 1-10 nA (2-probe constant voltage 1-5 $\mu$V) ac techniques for Hall bar (Corbino) devices.

\vspace{0.5cm}
\textbf{Acknowledgements}\\
Authors acknowledge support by the Department of Energy Awards DE-SC0008630 (T.W. and L.P.R.) and  DE-SC0010544 (Y.L-G and J.L.); by the Department of Defense Office of Naval Research Award N000141410339 (A.K., G.S., Y.L-G and L.P.R.);  by the National Science Foundation grant DMR-1610139 (Z.W.) and by the Gordon and Betty Moore Foundation (K.W., K.B. and L.N.P.).

\textbf{Authors contribution}\\
T.W., A.K. and Z.W. fabricated devices and performed measurements, G.S., J.L. and Y.L-G. developed theory and performed numerical simulations, K.W.W, K.B. and L.N.P. performed MBE growth and initial wafer characterization. A.K., L.P.R. and Y.L.-G. have written the paper with input from other authors. L.P.R. conceived and managed the project.

\vspace{0.5cm}
\textbf{Additional information}\\
Correspondence should be addressed to L.P.R. leonid@purdue.edu (experiment) or Y.L.-G. yuli@purdue.edu (theory).

\vspace{0.5cm}
\textbf{Competing financial interests}\\
The authors declare no competing financial interests.

%


\clearpage
\newpage
\onecolumngrid

\renewcommand{\thefigure}{S\arabic{figure}}
\renewcommand{\theequation}{S\arabic{equation}}
\renewcommand{\thetable}{S\arabic{table}}
\renewcommand{\thepage}{sup-\arabic{page}}
\setcounter{page}{1}
\setcounter{equation}{0}
\setcounter{figure}{0}
\setcounter{table}{0}

\begin{center}
\textbf{\Large Supplementary Materials} \\
\vspace{0.2in} \textsc{Parafermion supporting platform based on spin transitions in the fractional quantum Hall effect regime}\\
{\it Tailung Wu, Aleksandr Kazakov, George Simion, Zhong Wan, Jingcheng Liang, Kenneth W. West, Kirk Baldwin, Loren N. Pfeiffer, Yuli Lyanda-Geller, and Leonid P. Rokhinson}
\end{center}

\subsection{Comparison of spin transition in different wafers}

\begin{figure}[h]
\centering\includegraphics[width=0.7\columnwidth]{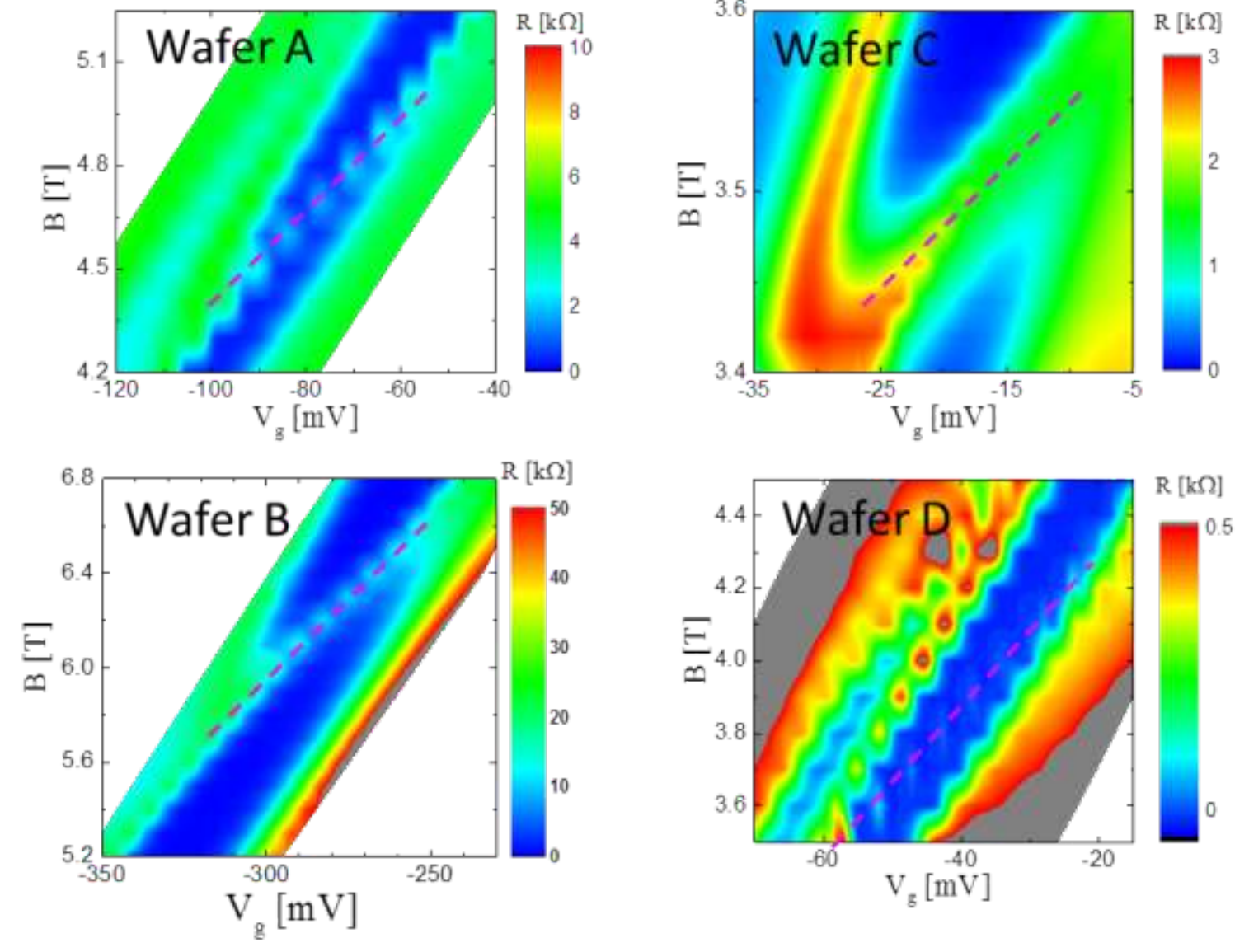}
\vspace{0in}
\caption{Filling factor $\nu=2/3$ spin transition measured in single gate Hall bar samples in different wafers. Dashed lines mark spin transitions.}
\label{fs:fig1}
\vspace{0in}
\end{figure}

Parameters of four wafers used in this study are summarized in Table~\ref{ts-wafers} and gate control of ferromagnetic spin transition in these wafers is demonstrated in Fig.~\ref{fs:fig1}.

\begin{table}[h]
\begin{tabular}{|c||l|l|l|l|}
  \hline \hline
  wafer & $d_1$ [nm] & $d_2$ [nm] & $n\times10^{11}$ [cm$^{-2}$] & $\mu\times 10^6$ [cm$^2$/Vs] \\  \hline \hline
  A & 135 & 110 & 1.3 & 3.8 \\  \hline
  B & 135 & 70 & 1.6 & 4.1 \\ \hline
  C & 185 & 270 & 0.8 & 3.7 \\ \hline
  D & 135 & 160 & 0.8 & 5.0 \\ \hline
  \hline

\end{tabular}
\caption{Parameters of several wafers used in the study. Distance $d_1$ is the thickness of GaAs layer, and $d_2$ is the $\delta$-doping setback below the heterointerface.}
\label{ts-wafers}
\end{table}

\clearpage
\subsection{Modeling of domain walls in fractional quantum Hall ferromagnets}

Quantum Hall ferromagnets exhibit a phase transition between spin-polarized and spin-unpolarized states \cite{Giuliani-book}.
It was shown in \cite{Kazakov2017a-sup, Simion2017-sup} that using electrostatic gates that make exchange interactions in a system position-dependent, it is possible to create both polarized and unpolarized state in a single sample with domain wall separating regions with different spin polarization. In Ref~[\onlinecite{Simion2017-sup}], an analytic theory of domain walls  for integer quantum hall effect ferromagnets was presented. It was further demonstarted that these domain walls lead to topological superconductivity when proximity--coupled to an s-type superconductor, with Majorana fermions forming at the boundary between trivial and topological superconducting phases.

  In investigation of fractional quantum Hall effect, edge states at the boundaries of the quantum Hall system and edge modes
flowing around antidots and constrictions has been intensively studied over the years \cite{macdonald_edge_1990,beenakker_edge_1990,wen_topological_1991,wen_theory_1992,chamon_sharp_1994}. In particular, at a fractional filling factor $\nu=2/3$, edge modes have been studied in both polarized and various kinds of unpolarized phases \cite{Jainedges,PhysRevB.51.9895,mcdonald_topological_1996,moore_edge_1997}.
However, microscopic consideration of the boundary between polarized and unpolarized phases presents a challenge.
Here we demonstrate emergence of a domain wall between polarized and unpolarized states in fractional quantum Hall ferromagnets by using a method of exact diagonalization in a system with small number of particles.

In order to understand the range of parameters we first obtain the quantum phase transition  between spin-polarized and spin-unpolarized states using exact diagonalization on the Haldane sphere with a monopole charge 2Q producing a magnetic field\cite{haldane83-sup}. The Haldane pseudopotentials, characterizing the interactions on the sphere, are defined by electron wavefunctions in the triangular quantum well and depend on
the electric field. For $ N$ particles on the sphere at a filling gactor $\nu=2/3$, we obtain that the polarized state emerges at $2Q=3/2N$ and unpolarized state arises at $2Q=3/2N-1$. As these monopoles  are different, we introduce a regularization procedure yielding the difference between energies of polarized and unpolarized states $\Delta=(E_p-E_u) \frac{3/2 N} {3/2N-1}$, where $E_p$ and $E_u$ are ground state energies for $2Q=3/2N$ and  $2Q=3/2N-1$, correspondingly. For  $\Delta>0$ the unpolarized state is the ground state. We find that an electric field of $4.8\times 10^{4}$ V/cm induces the spin polarization transition.

For creating reconfigurable network of helical domain walls in fractional quantum Hall ferromagnets, it is crucial to understand
physics of edge states generated by electrostatic gates or in the presence of a varying g-factor. In previous works edge states have been studied numerically only in ideal fractional quantum Hall systems near sample boundaries \cite{Jainedges}, but not at the boundary between polarized and unpolarized phases. The Haldane sphere cannot be used for simulation of such a boundary because spin-polarized and spin-unpolarized states appear at different monopole strengths. Investigation of edge states and their control using gates or inhomogeneous g-factors can be numerically studied using disk geometry \cite{Rezayidisk,TylanTyler2017}.

We simulate a system of electrons in a magnetic field in a quantum Hall ferromagnet on the disk. Long range Coulomb interactions among electrons
are introduced using Haldane pseudopotentials. A neutralizing background and a confinement potential are used to hold the
electrons together.   We use a parabolic
confinement $U(r)=C r^2$, where $C=0.036 e^2/\epsilon l_m^3$.  A realistic
confinement involves placing a neutralizing background close to the 2D
electron gas. However, Laughlin states are observed only if the distance  between
the electron gas and the background is smaller than one magnetic length,
which is much smaller than the characteristic separation of the doping layer from the 2D gas in experiments. Small separation of background charge from the disk is a limitation of the disk model.

 In our modeling we use spatially-dependent Zeeman effect to control the spin polarization of the 2DEG. As is evident from experiments, spin transition can be controlled by modulation of either Coulomb or Zeeman energies interchangeably, see e.g. Fig.~\ref{f:fig3}. Introducing spatially-dependent Zeeman splitting has computational advantages. In our model, the central region of the disk of radius $R_1=2.9 l_m$ is characterized by a large Zeeman  term $E_Z^{max}$, while the outer region with the outer diameter $R_2=4.8 l_m$ is set to $E_Z^{min}=0$. The Zeeman term is varied smoothly within the region
$R_1< r< R_1+\Delta R$, where $\Delta R=0.4 l_m$, resulting in a smooth variation of wavefunctions across the disk and avoiding spurious effects originating from abrupt changes. Note that due to a strong penetration of electron wavefunctions from the outer $R_1<r<R_2$ region into the inner $r<R_1$ region the difference of the average spin splitting $\int\psi(r)^*E_Z(r)\psi(r)d^2r$ for the modes on the two sides of the domain wall is $<6\%$, similar to the experimental conditions. Therefore, our model reflects soft edge characterizing the experiment.

 The total electron Hamiltonian is given by
\begin{eqnarray}
&{\cal H}=\frac{1}{2m^*}\sum_i (\mathbf{p}+\frac{e\mathbf{A}}{c})_i^2 + E_z(r_i)\sigma_z^{(i)}+U_i\nonumber\\
&+ \sum_{ij}\frac{e^2}{\epsilon|\mathbf{r}_i-\mathbf{r}_j|}
\end{eqnarray}
The Hamiltonian is diagonalized using a
configuration interaction method. The states are classified
according to their (conserved) projection of total angular momentum
on $z$-axis, $L_z$, and the total spin of electrons. The main challenge is the computational complexity of spin unpolarized states,
as their Hilbert space is significantly increased by the presence of both spin
orientations of electrons, and the system that includes both spin-polarized and spin-unpolarized regions.

We place $N$ electrons on a disk, allowing only single particle
states with $0\leq m \leq 3 N/2  -1$, resulting in $R_2=\sqrt{3N-2}l_m$. Only states from the lowest Landau
level $n=0$ with spin up and spin down polarizations are included. The region of polarized states
is characterized by a radius $R_1=l_m\sqrt{3 N/2-1}$.
The typical density distribution has two contributions: from the spin polarized region in the interior of the disk and from the spin-unpolarized region in the exterior of the disk. The electron density for $r>R_1$ is composed
of the tails of the wavefunctions and is a manifestation of soft
confinement of charge carriers that we choose. Spin polarization varies from $1$ inside the disk to zero closer to $R_2$.

We have included up to 12 electrons in the exact diagonalization calculation for the polarized state at a filling factor $2/3$,
and 8 electrons for unpolarized phases and modeling of  polarized and unpolarized phases in a disk with spatially dependent Zeeman term. We identified the ground state, which is spin-polarized in the center and unpolarized in the outer region of the disk, as well as the edge states flowing close to the boundary between spin-polarized and spin-unpolarized regions. Energies and wavefunctions for edge states, their electron and spin density distributions are calculated. For 8 particles, the ground state corresponds to the total angular momentum projection $ L_z=46$ and total spin of 8 particles $S_z=2$.
That is, ground state of the disk with inhomogeneous Zeeman splitting is given by $3N/4$ electrons in the spin-up states
and $N/4$ electrons in the spin-down states. In the composite
fermion transformation, two vortices are attached to each electron
and they completely fill the lowest composite fermion $\Lambda$-level with spin up
($0\leq m \leq N/2-1$) and partially fill two composite fermion $\Lambda$-levels with spin up ($-1\leq
m \leq N/2-2$) and spin down ($N/2\leq m\leq N-1$). The composite fermion transformation
defines the angular momentum of the ground state:
\begin{equation}
L_z=L^{CF}_z+pN(N-1)=-\frac{N(N-3)}{4} +N(N-1)=\frac{N(3N-1)}{4}.
\end{equation}
For $N=8$, $ L_z=46$ indeed, coinciding with the result of our numerical simulation. The ground state is separated by the gap from the rest of the spectra, Fig.~\ref{FigSpectra}, and does not carry current. Ground state is spin polarized in the interior part of the disk and spin-unpolarized in the exterior area, as expected Fig.~\ref{Ground}.

The lowest energy excitations which have spin polarization of the ground state and correspond to a single flux addition or subtraction from the ground state have $L_z=45$ and $L_z=47$, see Fig.~\ref{FigSpectra}.  These are the modes that carry electrical current. In a disk geometry these two modes have different angular velocities. When mapped onto a plane, these states will have different linear velocities, i.e. have counter-propagating components. In Fig.~\ref{FigDiff}, we show the results for the difference of spin densities of the two modes near the domain wall between polarized and unpolarized region. Despite finite size effects in a small system, the exact diagonalization allows clearly identify that the two edge states in the domain wall region have components of spin density with opposite orientation.
Thus, having opposite components of velocity and spin, these states can potentially be coupled to an s-type superconductor, a pre-requisite for generating topological superconductivity. In integer quantum Hall ferromagnets \cite{Simion2017-sup} proximity superconducting coupling has resulted in topological supercondutivity in the domain wall region and Majorana zero modes at the boundaries between superconducting regions with non-trivial topological and trivial s-type order. In the fractional quantum Hall ferromagnet, because of the difference in the degeneracy of composite fermion $\Lambda$-levels, the parafermion modes are expected to arise at the boundary of topological and non-topological superconductor\cite{Clarke2012-sup}.
\begin{figure}
\centering
\includegraphics[width=0.8\textwidth]{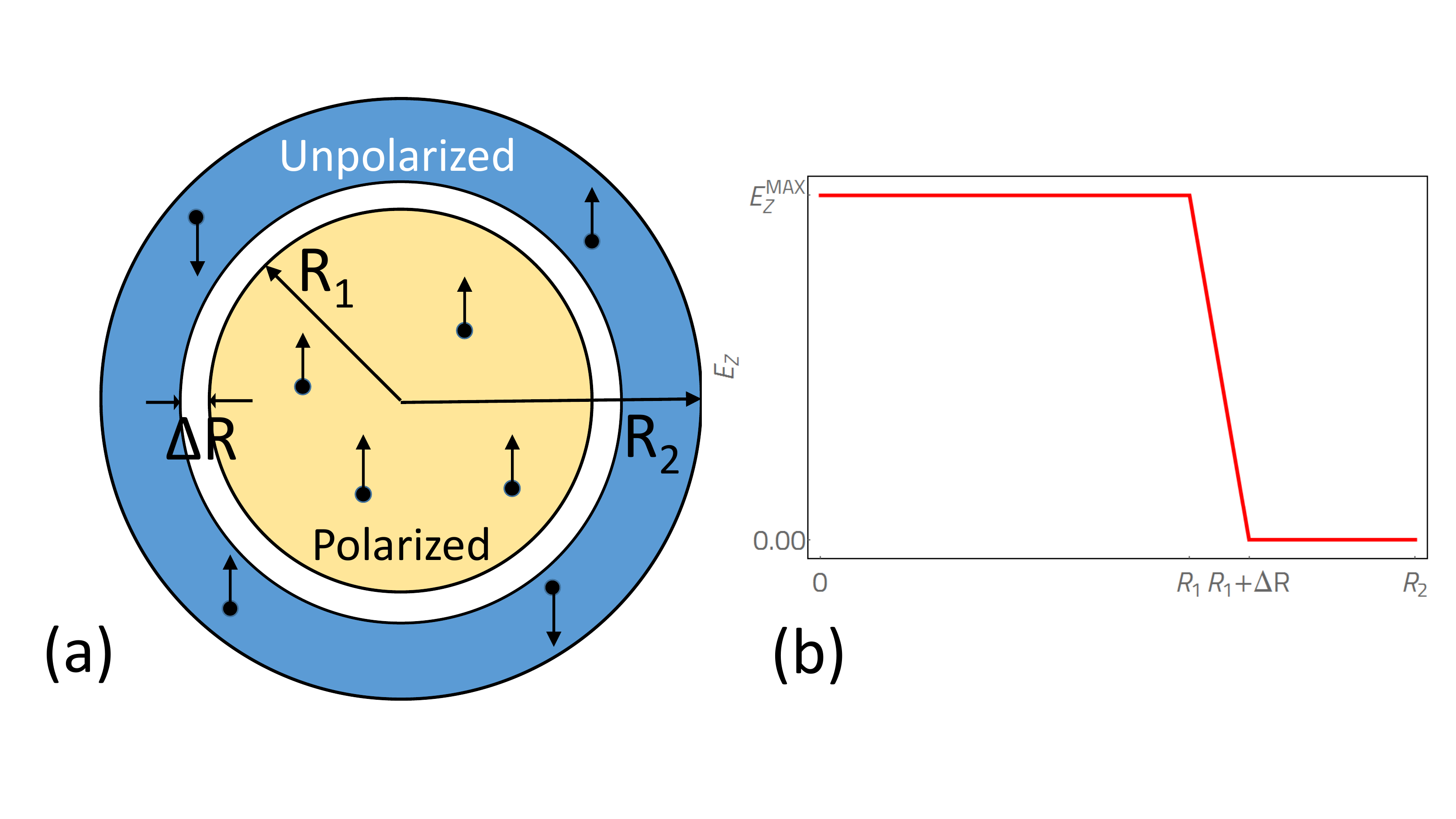}
\caption{Model for simulation of edge states near electrostatically induced boundary; (a) disk with three regions; ( b) profile of Zeemann interactions.}
\label{simulationmodel}
\end{figure}

\begin{figure}
\centering
\includegraphics[width=0.8\textwidth]{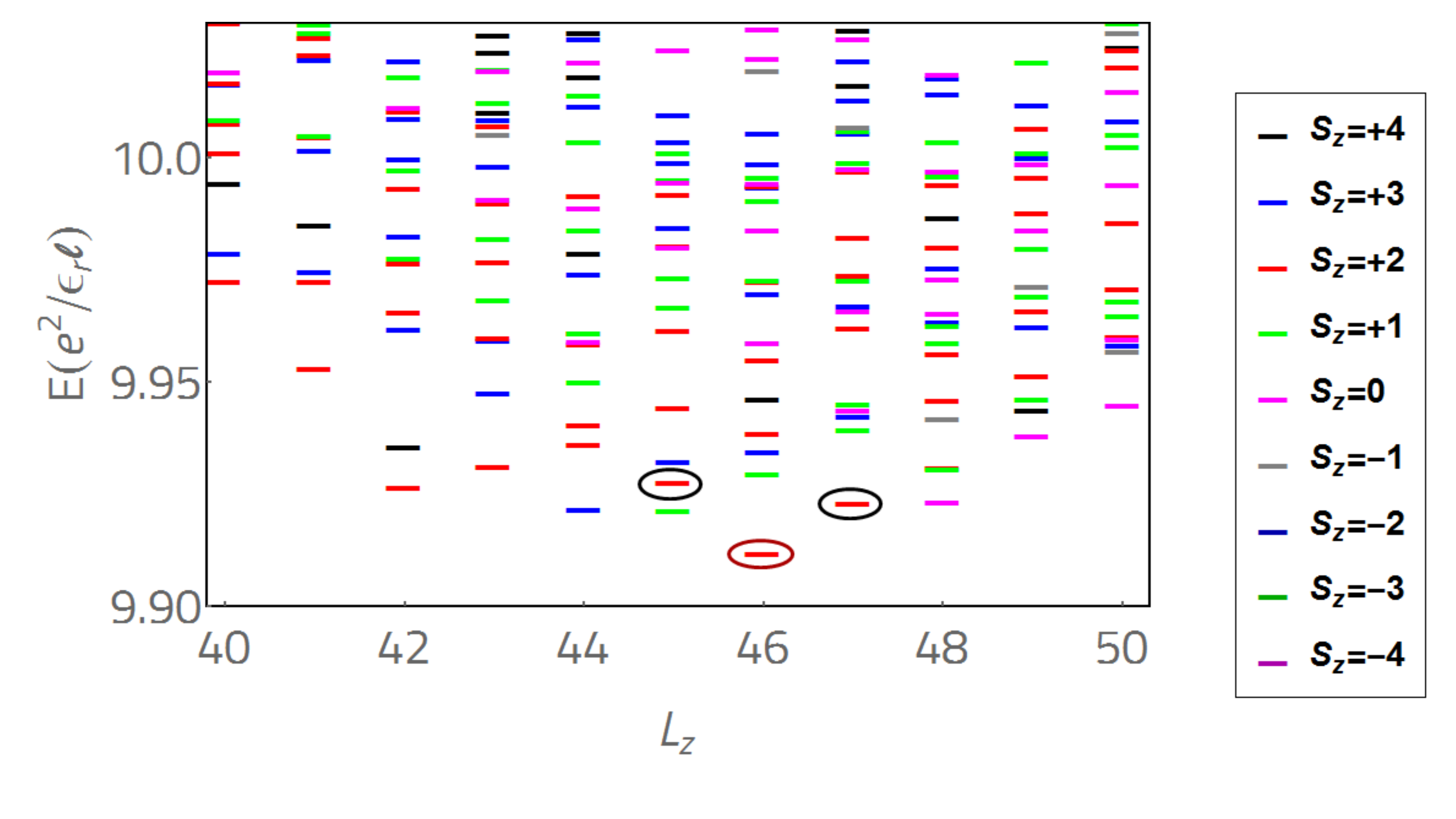},
\caption{Spectra of 8 electrons on the disk with profile of Zeeman interactions shown in Fig.\ref{simulationmodel}. The spectra are characterized by total angular momentum $L_z$ and total spin of particles $S_z$. Ground state at $L_z=46$ and $S_z=2$
is circled red. Edge excitations with the same $S_z=2$ as in the ground state and with $L_z=45, 47$, which correspond to the addition or subtraction of a single flux,  are circled black.}
\label{FigSpectra}
\end{figure}

\begin{figure}
\centering
\includegraphics[width=0.7\textwidth]{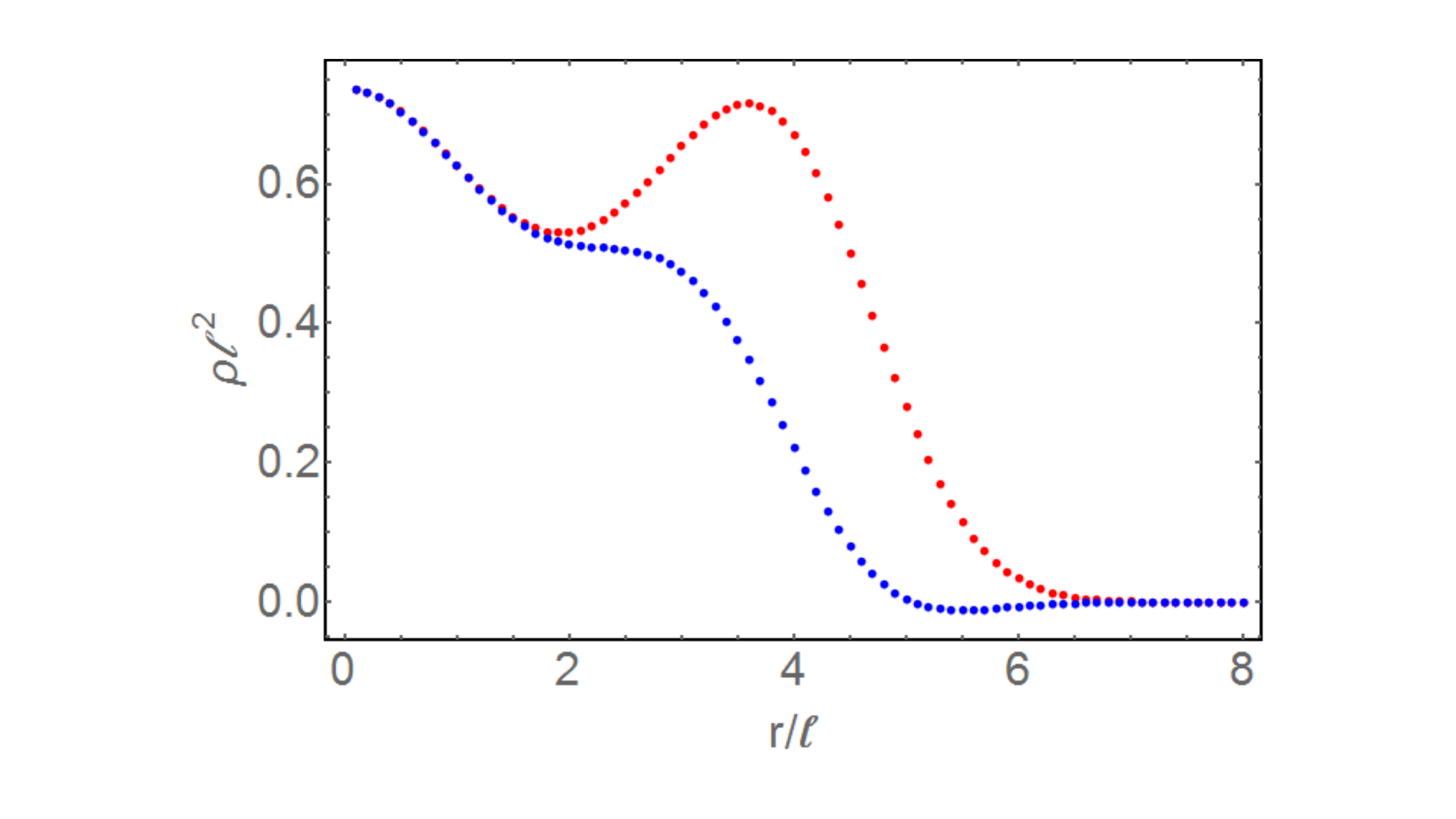},
\caption{The ground state electron density (red) and spin density (blue) for 8 electrons on a disk containing the domain wall between polarized and  unpolarized states at a filling factor 2/3 in a magnetic field.}
\label{Ground}
\end{figure}

\begin{figure}
\centering
\includegraphics[width=0.7\textwidth]{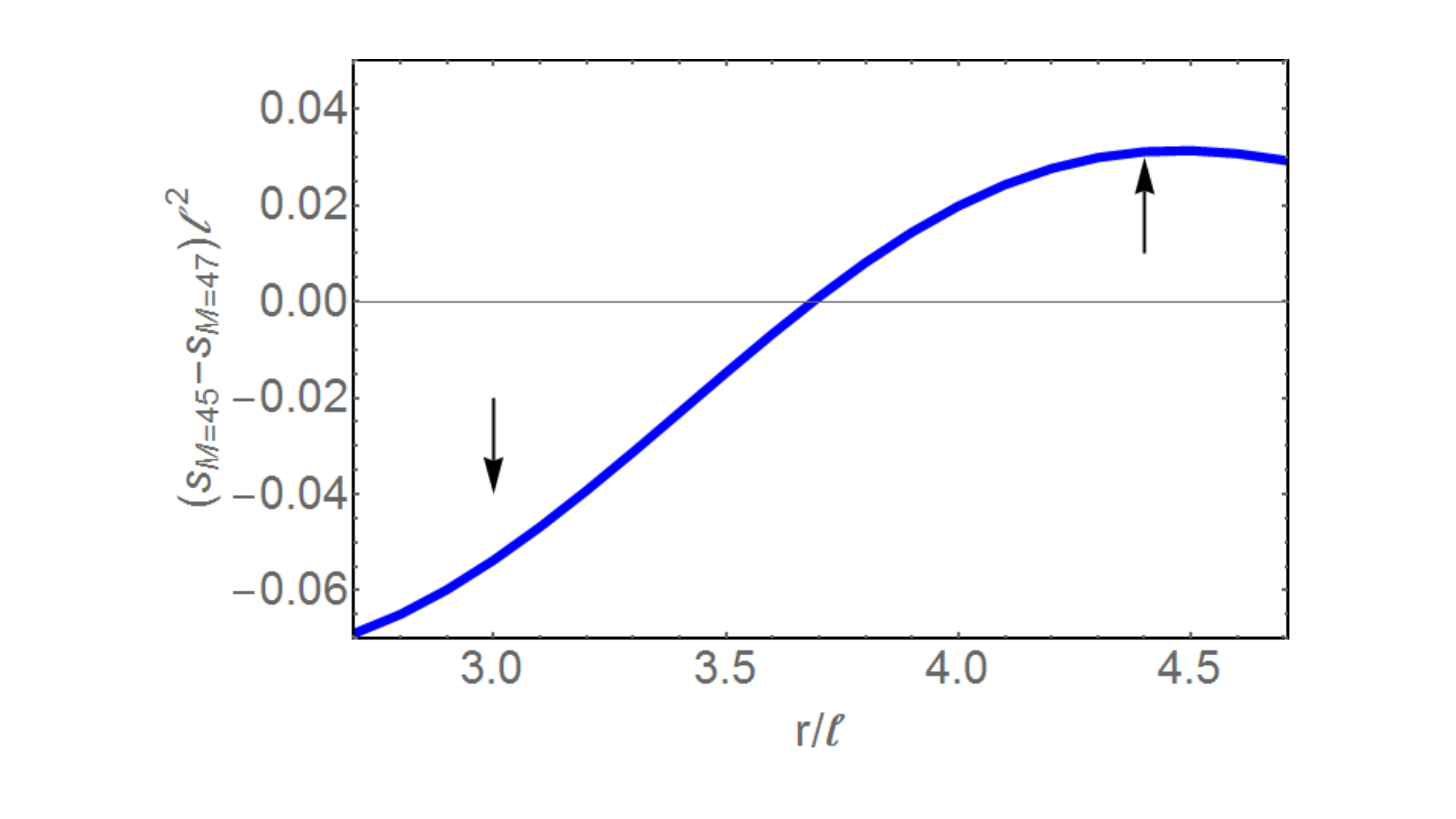},
\caption{Difference of spin densities of the two edge states with $ L_z=45$ and $L_z=47$ in the region of the disk  containing the domain wall between polarized and  unpolarized states at a filling factor 2/3 in a magnetic field.}
\label{FigDiff}
\end{figure}


\begin{thebibliography}{24}%
\makeatletter
\providecommand \@ifxundefined [1]{%
 \@ifx{#1\undefined}
}%
\providecommand \@ifnum [1]{%
 \ifnum #1\expandafter \@firstoftwo
 \else \expandafter \@secondoftwo
 \fi
}%
\providecommand \@ifx [1]{%
 \ifx #1\expandafter \@firstoftwo
 \else \expandafter \@secondoftwo
 \fi
}%
\providecommand \natexlab [1]{#1}%
\providecommand \enquote  [1]{``#1''}%
\providecommand \bibnamefont  [1]{#1}%
\providecommand \bibfnamefont [1]{#1}%
\providecommand \citenamefont [1]{#1}%
\providecommand \href@noop [0]{\@secondoftwo}%
\providecommand \href [0]{\begingroup \@sanitize@url \@href}%
\providecommand \@href[1]{\@@startlink{#1}\@@href}%
\providecommand \@@href[1]{\endgroup#1\@@endlink}%
\providecommand \@sanitize@url [0]{\catcode `\\12\catcode `\$12\catcode
  `\&12\catcode `\#12\catcode `\^12\catcode `\_12\catcode `\%12\relax}%
\providecommand \@@startlink[1]{}%
\providecommand \@@endlink[0]{}%
\providecommand \url  [0]{\begingroup\@sanitize@url \@url }%
\providecommand \@url [1]{\endgroup\@href {#1}{\urlprefix }}%
\providecommand \urlprefix  [0]{URL }%
\providecommand \Eprint [0]{\href }%
\providecommand \doibase [0]{http://dx.doi.org/}%
\providecommand \selectlanguage [0]{\@gobble}%
\providecommand \bibinfo  [0]{\@secondoftwo}%
\providecommand \bibfield  [0]{\@secondoftwo}%
\providecommand \translation [1]{[#1]}%
\providecommand \BibitemOpen [0]{}%
\providecommand \bibitemStop [0]{}%
\providecommand \bibitemNoStop [0]{.\EOS\space}%
\providecommand \EOS [0]{\spacefactor3000\relax}%
\providecommand \BibitemShut  [1]{\csname bibitem#1\endcsname}%
\let\auto@bib@innerbib\@empty
\bibitem [{\citenamefont {Nayak}\ \emph {et~al.}(2008)\citenamefont {Nayak},
  \citenamefont {Simon}, \citenamefont {Stern}, \citenamefont {Freedman},\ and\
  \citenamefont {Sarma}}]{Nayak2008}%
  \BibitemOpen
  \bibfield  {author} {\bibinfo {author} {\bibfnamefont {Chetan}\ \bibnamefont
  {Nayak}}, \bibinfo {author} {\bibfnamefont {Steven~H.}\ \bibnamefont
  {Simon}}, \bibinfo {author} {\bibfnamefont {Ady}\ \bibnamefont {Stern}},
  \bibinfo {author} {\bibfnamefont {Michael}\ \bibnamefont {Freedman}}, \ and\
  \bibinfo {author} {\bibfnamefont {Sankar~Das}\ \bibnamefont {Sarma}},\
  }\bibfield  {title} {\enquote {\bibinfo {title} {Non-abelian anyons and
  topological quantum computation},}\ }\href@noop {} {\bibfield  {journal}
  {\bibinfo  {journal} {Rev. Mod. Phys.}\ }\textbf {\bibinfo {volume} {80}},\
  \bibinfo {pages} {1083 --1159} (\bibinfo {year} {2008})}\BibitemShut
  {NoStop}%
\bibitem [{\citenamefont {Baraban}\ \emph {et~al.}(2010)\citenamefont
  {Baraban}, \citenamefont {Bonesteel},\ and\ \citenamefont
  {Simon}}]{Baraban2010}%
  \BibitemOpen
  \bibfield  {author} {\bibinfo {author} {\bibfnamefont {M.}~\bibnamefont
  {Baraban}}, \bibinfo {author} {\bibfnamefont {N.~E.}\ \bibnamefont
  {Bonesteel}}, \ and\ \bibinfo {author} {\bibfnamefont {S.~H.}\ \bibnamefont
  {Simon}},\ }\bibfield  {title} {\enquote {\bibinfo {title} {Resources
  required for topological quantum factoring},}\ }\href {\doibase
  10.1103/PhysRevA.81.062317} {\bibfield  {journal} {\bibinfo  {journal} {Phys.
  Rev. A}\ }\textbf {\bibinfo {volume} {81}},\ \bibinfo {pages} {062317}
  (\bibinfo {year} {2010})}\BibitemShut {NoStop}%
\bibitem [{\citenamefont {Fendley}(2012)}]{Fendley2012}%
  \BibitemOpen
  \bibfield  {author} {\bibinfo {author} {\bibfnamefont {Paul}\ \bibnamefont
  {Fendley}},\ }\bibfield  {title} {\enquote {\bibinfo {title} {Parafermionic
  edge zero modes in $z_n$-invariant spin chains},}\ }\href@noop {} {\bibfield
  {journal} {\bibinfo  {journal} {J. Stat. Mech.}\ ,\ \bibinfo {pages}
  {P11020}} (\bibinfo {year} {2012})}\BibitemShut {NoStop}%
\bibitem [{\citenamefont {Alicea}\ and\ \citenamefont
  {Fendley}(2016)}]{Alicea2016}%
  \BibitemOpen
  \bibfield  {author} {\bibinfo {author} {\bibfnamefont {Jason}\ \bibnamefont
  {Alicea}}\ and\ \bibinfo {author} {\bibfnamefont {Paul}\ \bibnamefont
  {Fendley}},\ }\bibfield  {title} {\enquote {\bibinfo {title} {Topological
  phases with parafermions: Theory and blueprints},}\ }\href {\doibase
  10.1146/annurev-conmatphys-031115-011336} {\bibfield  {journal} {\bibinfo
  {journal} {Annual Review of Condensed Matter Physics}\ }\textbf {\bibinfo
  {volume} {7}},\ \bibinfo {pages} {119--139} (\bibinfo {year}
  {2016})}\BibitemShut {NoStop}%
\bibitem [{\citenamefont {Mong}\ \emph {et~al.}(2014)\citenamefont {Mong},
  \citenamefont {Clarke}, \citenamefont {Alicea}, \citenamefont {Lindner},
  \citenamefont {Fendley}, \citenamefont {Nayak}, \citenamefont {Oreg},
  \citenamefont {Stern}, \citenamefont {Berg}, \citenamefont {Shtengel},\ and\
  \citenamefont {Fisher}}]{Mong2013}%
  \BibitemOpen
  \bibfield  {author} {\bibinfo {author} {\bibfnamefont {Roger S.~K.}\
  \bibnamefont {Mong}}, \bibinfo {author} {\bibfnamefont {David~J.}\
  \bibnamefont {Clarke}}, \bibinfo {author} {\bibfnamefont {Jason}\
  \bibnamefont {Alicea}}, \bibinfo {author} {\bibfnamefont {Netanel~H.}\
  \bibnamefont {Lindner}}, \bibinfo {author} {\bibfnamefont {Paul}\
  \bibnamefont {Fendley}}, \bibinfo {author} {\bibfnamefont {Chetan}\
  \bibnamefont {Nayak}}, \bibinfo {author} {\bibfnamefont {Yuval}\ \bibnamefont
  {Oreg}}, \bibinfo {author} {\bibfnamefont {Ady}\ \bibnamefont {Stern}},
  \bibinfo {author} {\bibfnamefont {Erez}\ \bibnamefont {Berg}}, \bibinfo
  {author} {\bibfnamefont {Kirill}\ \bibnamefont {Shtengel}}, \ and\ \bibinfo
  {author} {\bibfnamefont {Matthew P.~A.}\ \bibnamefont {Fisher}},\ }\bibfield
  {title} {\enquote {\bibinfo {title} {Universal topological quantum
  computation from a superconductor-{Abelian} quantum hall heterostructure},}\
  }\href {\doibase 10.1103/PhysRevX.4.011036} {\bibfield  {journal} {\bibinfo
  {journal} {Phys. Rev. X}\ }\textbf {\bibinfo {volume} {4}},\ \bibinfo {pages}
  {011036} (\bibinfo {year} {2014})}\BibitemShut {NoStop}%
\bibitem [{\citenamefont {Clarke}\ \emph {et~al.}(2012)\citenamefont {Clarke},
  \citenamefont {Alicea},\ and\ \citenamefont {Shtengel}}]{Clarke2012}%
  \BibitemOpen
  \bibfield  {author} {\bibinfo {author} {\bibfnamefont {David~J.}\
  \bibnamefont {Clarke}}, \bibinfo {author} {\bibfnamefont {Jason}\
  \bibnamefont {Alicea}}, \ and\ \bibinfo {author} {\bibfnamefont {Kirill}\
  \bibnamefont {Shtengel}},\ }\bibfield  {title} {\enquote {\bibinfo {title}
  {Exotic non-{A}belian anyons from conventional fractional quantum {H}all
  states},}\ }\href {\doibase doi:10.1038/ncomms2340} {\bibfield  {journal}
  {\bibinfo  {journal} {Nat. Commun.}\ }\textbf {\bibinfo {volume} {4}},\
  \bibinfo {pages} {1348} (\bibinfo {year} {2012})}\BibitemShut {NoStop}%
\bibitem [{\citenamefont {Maciejko}\ \emph {et~al.}(2011)\citenamefont
  {Maciejko}, \citenamefont {Hughes},\ and\ \citenamefont
  {Zhang}}]{Maciejko2011}%
  \BibitemOpen
  \bibfield  {author} {\bibinfo {author} {\bibfnamefont {Joseph}\ \bibnamefont
  {Maciejko}}, \bibinfo {author} {\bibfnamefont {Taylor~L.}\ \bibnamefont
  {Hughes}}, \ and\ \bibinfo {author} {\bibfnamefont {Shou-Cheng}\ \bibnamefont
  {Zhang}},\ }\bibfield  {title} {\enquote {\bibinfo {title} {The quantum spin
  Hall effect},}\ }\href {\doibase 10.1146/annurev-conmatphys-062910-140538}
  {\bibfield  {journal} {\bibinfo  {journal} {Annual Review of Condensed Matter
  Physics}\ }\textbf {\bibinfo {volume} {2}},\ \bibinfo {pages} {31--53}
  (\bibinfo {year} {2011})}\BibitemShut {NoStop}%
\bibitem [{\citenamefont {Hasan}\ and\ \citenamefont
  {Moore}(2011)}]{Hasan2011}%
  \BibitemOpen
  \bibfield  {author} {\bibinfo {author} {\bibfnamefont {M.~Zahid}\
  \bibnamefont {Hasan}}\ and\ \bibinfo {author} {\bibfnamefont {Joel~E.}\
  \bibnamefont {Moore}},\ }\bibfield  {title} {\enquote {\bibinfo {title}
  {Three-dimensional topological insulators},}\ }\href {\doibase
  10.1146/annurev-conmatphys-062910-140432} {\bibfield  {journal} {\bibinfo
  {journal} {Annual Review of Condensed Matter Physics}\ }\textbf {\bibinfo
  {volume} {2}},\ \bibinfo {pages} {55--78} (\bibinfo {year}
  {2011})}\BibitemShut {NoStop}%
\bibitem [{\citenamefont {Lutchyn}\ \emph {et~al.}(2010)\citenamefont
  {Lutchyn}, \citenamefont {Sau},\ and\ \citenamefont {Sarma}}]{Lutchyn2010a}%
  \BibitemOpen
  \bibfield  {author} {\bibinfo {author} {\bibfnamefont {Roman~M.}\
  \bibnamefont {Lutchyn}}, \bibinfo {author} {\bibfnamefont {Jay~D.}\
  \bibnamefont {Sau}}, \ and\ \bibinfo {author} {\bibfnamefont {S.~Das}\
  \bibnamefont {Sarma}},\ }\bibfield  {title} {\enquote {\bibinfo {title}
  {Majorana fermions and a topological phase transition in
  semiconductor-superconductor heterostructures},}\ }\href@noop {} {\bibfield
  {journal} {\bibinfo  {journal} {Phys. Rev. Lett.}\ }\textbf {\bibinfo
  {volume} {105}},\ \bibinfo {pages} {077001} (\bibinfo {year}
  {2010})}\BibitemShut {NoStop}%
\bibitem [{\citenamefont {Oreg}\ \emph {et~al.}(2010)\citenamefont {Oreg},
  \citenamefont {Refael},\ and\ \citenamefont {von Oppen}}]{Oreg2010}%
  \BibitemOpen
  \bibfield  {author} {\bibinfo {author} {\bibfnamefont {Yuval}\ \bibnamefont
  {Oreg}}, \bibinfo {author} {\bibfnamefont {Gil}\ \bibnamefont {Refael}}, \
  and\ \bibinfo {author} {\bibfnamefont {Felix}\ \bibnamefont {von Oppen}},\
  }\bibfield  {title} {\enquote {\bibinfo {title} {Helical liquids and
  {M}ajorana bound states in quantum wires},}\ }\href {\doibase
  10.1103/PhysRevLett.105.177002} {\bibfield  {journal} {\bibinfo  {journal}
  {Phys. Rev. Lett.}\ }\textbf {\bibinfo {volume} {105}},\ \bibinfo {pages}
  {177002} (\bibinfo {year} {2010})}\BibitemShut {NoStop}%
\bibitem [{\citenamefont {Fal'ko}\ and\ \citenamefont
  {Iordanskii}(2000)}]{falko00}%
  \BibitemOpen
  \bibfield  {author} {\bibinfo {author} {\bibfnamefont {V.I.}\ \bibnamefont
  {Fal'ko}}\ and\ \bibinfo {author} {\bibfnamefont {S.V.}\ \bibnamefont
  {Iordanskii}},\ }\bibfield  {title} {\enquote {\bibinfo {title} {Spin-orbit
  coupling effect on quantum {H}all ferromagnets with vanishing {Z}eeman
  energy},}\ }\href@noop {} {\bibfield  {journal} {\bibinfo  {journal} {Phys.
  Rev. Lett.}\ }\textbf {\bibinfo {volume} {84}},\ \bibinfo {pages} {127 -- 30}
  (\bibinfo {year} {2000})}\BibitemShut {NoStop}%
\bibitem [{\citenamefont {Kazakov}\ \emph {et~al.}(2017)\citenamefont
  {Kazakov}, \citenamefont {Simion}, \citenamefont {Lyanda-Geller},
  \citenamefont {Kolkovsky}, \citenamefont {Adamus}, \citenamefont
  {Karczewski}, \citenamefont {Wojtowicz},\ and\ \citenamefont
  {Rokhinson}}]{Kazakov2017a}%
  \BibitemOpen
  \bibfield  {author} {\bibinfo {author} {\bibfnamefont {Aleksandr}\
  \bibnamefont {Kazakov}}, \bibinfo {author} {\bibfnamefont {George}\
  \bibnamefont {Simion}}, \bibinfo {author} {\bibfnamefont {Yuli}\ \bibnamefont
  {Lyanda-Geller}}, \bibinfo {author} {\bibfnamefont {Valery}\ \bibnamefont
  {Kolkovsky}}, \bibinfo {author} {\bibfnamefont {Zbigniew}\ \bibnamefont
  {Adamus}}, \bibinfo {author} {\bibfnamefont {Grzegorz}\ \bibnamefont
  {Karczewski}}, \bibinfo {author} {\bibfnamefont {Tomasz}\ \bibnamefont
  {Wojtowicz}}, \ and\ \bibinfo {author} {\bibfnamefont {Leonid~P.}\
  \bibnamefont {Rokhinson}},\ }\bibfield  {title} {\enquote {\bibinfo {title}
  {Mesoscopic transport in electrostatically defined spin-full channels in
  quantum {H}all ferromagnets},}\ }\href {\doibase
  10.1103/PhysRevLett.119.046803} {\bibfield  {journal} {\bibinfo  {journal}
  {Phys. Rev. Lett.}\ }\textbf {\bibinfo {volume} {119}},\ \bibinfo {pages}
  {046803} (\bibinfo {year} {2017})}\BibitemShut {NoStop}%
\bibitem [{\citenamefont {Simion}\ \emph {et~al.}(2017)\citenamefont {Simion},
  \citenamefont {Kazakov}, \citenamefont {Rokhinson}, \citenamefont
  {Wojtowicz},\ and\ \citenamefont {Lyanda-Geller}}]{Simion2017}%
  \BibitemOpen
  \bibfield  {author} {\bibinfo {author} {\bibfnamefont {G.~E.}\ \bibnamefont
  {Simion}}, \bibinfo {author} {\bibfnamefont {A.}~\bibnamefont {Kazakov}},
  \bibinfo {author} {\bibfnamefont {L.~P.}\ \bibnamefont {Rokhinson}}, \bibinfo
  {author} {\bibfnamefont {T.}~\bibnamefont {Wojtowicz}}, \ and\ \bibinfo
  {author} {\bibfnamefont {Y.~B.}\ \bibnamefont {Lyanda-Geller}},\ }\bibfield
  {title} {\enquote {\bibinfo {title} {Disorder-generated non-abelions},}\
  }\href@noop {} {\  (\bibinfo {year} {2017})},\ \Eprint
  {http://arxiv.org/abs/1707.02929} {arXiv:1707.02929} \BibitemShut {NoStop}%
\bibitem [{\citenamefont {Eisenstein}\ \emph {et~al.}(1990)\citenamefont
  {Eisenstein}, \citenamefont {Stormer}, \citenamefont {Pfeiffer},\ and\
  \citenamefont {West}}]{Eisenstein1990}%
  \BibitemOpen
  \bibfield  {author} {\bibinfo {author} {\bibfnamefont {J.P.}\ \bibnamefont
  {Eisenstein}}, \bibinfo {author} {\bibfnamefont {H.L.}\ \bibnamefont
  {Stormer}}, \bibinfo {author} {\bibfnamefont {L.N.}\ \bibnamefont
  {Pfeiffer}}, \ and\ \bibinfo {author} {\bibfnamefont {K.W.}\ \bibnamefont
  {West}},\ }\bibfield  {title} {\enquote {\bibinfo {title} {Evidence for a
  spin transition in the $\nu$=2/3 fractional quantum {H}all effect},}\ }\href
  {\doibase 10.1103/physrevb.41.7910} {\bibfield  {journal} {\bibinfo
  {journal} {Physical Review B}\ }\textbf {\bibinfo {volume} {41}},\ \bibinfo
  {pages} {7910--7913} (\bibinfo {year} {1990})}\BibitemShut {NoStop}%
\bibitem [{\citenamefont {Smet}\ \emph {et~al.}(2001)\citenamefont {Smet},
  \citenamefont {Deutschmann}, \citenamefont {Wegscheider}, \citenamefont
  {Abstreiter},\ and\ \citenamefont {von Klitzing}}]{Smet2001}%
  \BibitemOpen
  \bibfield  {author} {\bibinfo {author} {\bibfnamefont {J.~H.}\ \bibnamefont
  {Smet}}, \bibinfo {author} {\bibfnamefont {R.~A.}\ \bibnamefont
  {Deutschmann}}, \bibinfo {author} {\bibfnamefont {W.}~\bibnamefont
  {Wegscheider}}, \bibinfo {author} {\bibfnamefont {G.}~\bibnamefont
  {Abstreiter}}, \ and\ \bibinfo {author} {\bibfnamefont {K.}~\bibnamefont {von
  Klitzing}},\ }\bibfield  {title} {\enquote {\bibinfo {title} {Ising
  ferromagnetism and domain morphology in the fractional quantum {H}all
  regime},}\ }\href {\doibase 10.1103/PhysRevLett.86.2412} {\bibfield
  {journal} {\bibinfo  {journal} {Phys. Rev. Lett.}\ }\textbf {\bibinfo
  {volume} {86}},\ \bibinfo {pages} {2412--2415} (\bibinfo {year}
  {2001})}\BibitemShut {NoStop}%
\bibitem [{\citenamefont {Verdene}\ \emph {et~al.}(2007)\citenamefont
  {Verdene}, \citenamefont {Martin}, \citenamefont {Gamez}, \citenamefont
  {Smet}, \citenamefont {von Klitzing}, \citenamefont {Mahalu}, \citenamefont
  {Schuh}, \citenamefont {Abstreiter},\ and\ \citenamefont
  {Yacoby}}]{Verdene2007}%
  \BibitemOpen
  \bibfield  {author} {\bibinfo {author} {\bibfnamefont {Basile}\ \bibnamefont
  {Verdene}}, \bibinfo {author} {\bibfnamefont {Jens}\ \bibnamefont {Martin}},
  \bibinfo {author} {\bibfnamefont {Gerardo}\ \bibnamefont {Gamez}}, \bibinfo
  {author} {\bibfnamefont {Jurgen}\ \bibnamefont {Smet}}, \bibinfo {author}
  {\bibfnamefont {Klaus}\ \bibnamefont {von Klitzing}}, \bibinfo {author}
  {\bibfnamefont {Diana}\ \bibnamefont {Mahalu}}, \bibinfo {author}
  {\bibfnamefont {Dieter}\ \bibnamefont {Schuh}}, \bibinfo {author}
  {\bibfnamefont {Gerhard}\ \bibnamefont {Abstreiter}}, \ and\ \bibinfo
  {author} {\bibfnamefont {Amir}\ \bibnamefont {Yacoby}},\ }\bibfield  {title}
  {\enquote {\bibinfo {title} {Microscopic manifestation of the spin phase
  transition at filling factor 2/3},}\ }\href {\doibase 10.1038/nphys588}
  {\bibfield  {journal} {\bibinfo  {journal} {Nat Phys}\ }\textbf {\bibinfo
  {volume} {3}},\ \bibinfo {pages} {392--396} (\bibinfo {year}
  {2007})}\BibitemShut {NoStop}%
\bibitem [{\citenamefont {Hayakawa}\ \emph {et~al.}(2012)\citenamefont
  {Hayakawa}, \citenamefont {Muraki},\ and\ \citenamefont
  {Yusa}}]{Hayakawa2012}%
  \BibitemOpen
  \bibfield  {author} {\bibinfo {author} {\bibfnamefont {Junichiro}\
  \bibnamefont {Hayakawa}}, \bibinfo {author} {\bibfnamefont {Koji}\
  \bibnamefont {Muraki}}, \ and\ \bibinfo {author} {\bibfnamefont
  {Go}~\bibnamefont {Yusa}},\ }\bibfield  {title} {\enquote {\bibinfo {title}
  {Real-space imaging of fractional quantum {H}all liquids},}\ }\href {\doibase
  10.1038/nnano.2012.209} {\bibfield  {journal} {\bibinfo  {journal} {Nature
  Nanotechnology}\ }\textbf {\bibinfo {volume} {8}},\ \bibinfo {pages} {31--35}
  (\bibinfo {year} {2012})}\BibitemShut {NoStop}%
\bibitem [{\citenamefont {Kane}\ \emph {et~al.}(1994)\citenamefont {Kane},
  \citenamefont {Fisher},\ and\ \citenamefont {Polchinski}}]{Kane1994}%
  \BibitemOpen
  \bibfield  {author} {\bibinfo {author} {\bibfnamefont {C.~L.}\ \bibnamefont
  {Kane}}, \bibinfo {author} {\bibfnamefont {Matthew P.~A.}\ \bibnamefont
  {Fisher}}, \ and\ \bibinfo {author} {\bibfnamefont {J.}~\bibnamefont
  {Polchinski}},\ }\bibfield  {title} {\enquote {\bibinfo {title} {Randomness
  at the edge: Theory of quantum {H}all transport at filling $\nu=2/3$},}\
  }\href {\doibase 10.1103/PhysRevLett.72.4129} {\bibfield  {journal} {\bibinfo
   {journal} {Phys. Rev. Lett.}\ }\textbf {\bibinfo {volume} {72}},\ \bibinfo
  {pages} {4129--4132} (\bibinfo {year} {1994})}\BibitemShut {NoStop}%
\bibitem [{\citenamefont {Bid}\ \emph {et~al.}(2010)\citenamefont {Bid},
  \citenamefont {Ofek}, \citenamefont {Inoue}, \citenamefont {Heiblum},
  \citenamefont {Kane}, \citenamefont {Umansky},\ and\ \citenamefont
  {Mahalu}}]{Bid2010}%
  \BibitemOpen
  \bibfield  {author} {\bibinfo {author} {\bibfnamefont {Aveek}\ \bibnamefont
  {Bid}}, \bibinfo {author} {\bibfnamefont {N.}~\bibnamefont {Ofek}}, \bibinfo
  {author} {\bibfnamefont {H.}~\bibnamefont {Inoue}}, \bibinfo {author}
  {\bibfnamefont {M.}~\bibnamefont {Heiblum}}, \bibinfo {author} {\bibfnamefont
  {C.~L.}\ \bibnamefont {Kane}}, \bibinfo {author} {\bibfnamefont
  {V.}~\bibnamefont {Umansky}}, \ and\ \bibinfo {author} {\bibfnamefont
  {D.}~\bibnamefont {Mahalu}},\ }\bibfield  {title} {\enquote {\bibinfo {title}
  {Observation of neutral modes in the fractional quantum {H}all regime},}\
  }\href {\doibase 10.1038/nature09277} {\bibfield  {journal} {\bibinfo
  {journal} {Nature}\ }\textbf {\bibinfo {volume} {466}},\ \bibinfo {pages}
  {585--590} (\bibinfo {year} {2010})}\BibitemShut {NoStop}%
\bibitem [{\citenamefont {Venkatachalam}\ \emph {et~al.}(2012)\citenamefont
  {Venkatachalam}, \citenamefont {Hart}, \citenamefont {Pfeiffer},
  \citenamefont {West},\ and\ \citenamefont {Yacoby}}]{Venkatachalam2012}%
  \BibitemOpen
  \bibfield  {author} {\bibinfo {author} {\bibfnamefont {Vivek}\ \bibnamefont
  {Venkatachalam}}, \bibinfo {author} {\bibfnamefont {Sean}\ \bibnamefont
  {Hart}}, \bibinfo {author} {\bibfnamefont {Loren}\ \bibnamefont {Pfeiffer}},
  \bibinfo {author} {\bibfnamefont {Ken}\ \bibnamefont {West}}, \ and\ \bibinfo
  {author} {\bibfnamefont {Amir}\ \bibnamefont {Yacoby}},\ }\bibfield  {title}
  {\enquote {\bibinfo {title} {Local thermometry of neutral modes on the
  quantum {H}all edge},}\ }\href {\doibase 10.1038/nphys2384} {\bibfield
  {journal} {\bibinfo  {journal} {Nat Phys}\ }\textbf {\bibinfo {volume} {8}},\
  \bibinfo {pages} {676--681} (\bibinfo {year} {2012})}\BibitemShut {NoStop}%
\bibitem [{\citenamefont {Jain}(2007)}]{JainCFbook2007}%
  \BibitemOpen
  \bibfield  {author} {\bibinfo {author} {\bibfnamefont {Jainendra~K}\
  \bibnamefont {Jain}},\ }\href@noop {} {\emph {\bibinfo {title} {Composite
  fermions}}}\ (\bibinfo  {publisher} {Cambridge University Press},\ \bibinfo
  {address} {Cambridge},\ \bibinfo {year} {2007})\BibitemShut {NoStop}%
\bibitem [{\citenamefont {Hu}\ \emph {et~al.}(2009)\citenamefont {Hu},
  \citenamefont {Rezayi}, \citenamefont {Wan},\ and\ \citenamefont
  {Yang}}]{Rezayidisk}%
  \BibitemOpen
  \bibfield  {author} {\bibinfo {author} {\bibfnamefont {Zi-Xiang}\
  \bibnamefont {Hu}}, \bibinfo {author} {\bibfnamefont {E.~H.}\ \bibnamefont
  {Rezayi}}, \bibinfo {author} {\bibfnamefont {Xin}\ \bibnamefont {Wan}}, \
  and\ \bibinfo {author} {\bibfnamefont {Kun}\ \bibnamefont {Yang}},\
  }\bibfield  {title} {\enquote {\bibinfo {title} {Edge-mode velocities and
  thermal coherence of quantum hall interferometers},}\ }\href@noop {}
  {\bibfield  {journal} {\bibinfo  {journal} {Phys. Rev. B}\ }\textbf {\bibinfo
  {volume} {80}},\ \bibinfo {pages} {235330} (\bibinfo {year}
  {2009})}\BibitemShut {NoStop}%
\bibitem [{\citenamefont {Tylan-Tyler}\ and\ \citenamefont
  {Lyanda-Geller}(2017)}]{TylanTyler2017}%
  \BibitemOpen
  \bibfield  {author} {\bibinfo {author} {\bibfnamefont {A.}~\bibnamefont
  {Tylan-Tyler}}\ and\ \bibinfo {author} {\bibfnamefont {Y.B.}\ \bibnamefont
  {Lyanda-Geller}},\ }\bibfield  {title} {\enquote {\bibinfo {title} {In-plane
  electric fields and the $\nu=5/2$ fractional quantum Hall effect in disc
  geometry},}\ }\href@noop {} {\bibfield  {journal} {\bibinfo  {journal} {Phys.
  Rev. B}\ }\textbf {\bibinfo {volume} {95}},\ \bibinfo {pages} {121302}
  (\bibinfo {year} {2017})}\BibitemShut {NoStop}%
\bibitem [{\citenamefont {Wan}\ \emph {et~al.}(2015)\citenamefont {Wan},
  \citenamefont {Kazakov}, \citenamefont {Manfra}, \citenamefont {Pfeiffer},
  \citenamefont {West},\ and\ \citenamefont {Rokhinson}}]{Wan2015}%
  \BibitemOpen
  \bibfield  {author} {\bibinfo {author} {\bibfnamefont {Zhong}\ \bibnamefont
  {Wan}}, \bibinfo {author} {\bibfnamefont {Aleksandr}\ \bibnamefont
  {Kazakov}}, \bibinfo {author} {\bibfnamefont {Michael~J.}\ \bibnamefont
  {Manfra}}, \bibinfo {author} {\bibfnamefont {Loren~N.}\ \bibnamefont
  {Pfeiffer}}, \bibinfo {author} {\bibfnamefont {Ken~W.}\ \bibnamefont {West}},
  \ and\ \bibinfo {author} {\bibfnamefont {Leonid~P.}\ \bibnamefont
  {Rokhinson}},\ }\bibfield  {title} {\enquote {\bibinfo {title} {Induced
  superconductivity in high-mobility two-dimensional electron gas in gallium
  arsenide heterostructures},}\ }\href {\doibase 10.1038/ncomms8426} {\bibfield
   {journal} {\bibinfo  {journal} {Nat Commun}\ }\textbf {\bibinfo {volume}
  {6}},\ \bibinfo {pages} {7426} (\bibinfo {year} {2015})}\BibitemShut
  {NoStop}%
\end{thebibliography}

\clearpage


\begin{thebibliography}{14}%
\makeatletter
\providecommand \@ifxundefined [1]{%
 \@ifx{#1\undefined}
}%
\providecommand \@ifnum [1]{%
 \ifnum #1\expandafter \@firstoftwo
 \else \expandafter \@secondoftwo
 \fi
}%
\providecommand \@ifx [1]{%
 \ifx #1\expandafter \@firstoftwo
 \else \expandafter \@secondoftwo
 \fi
}%
\providecommand \natexlab [1]{#1}%
\providecommand \enquote  [1]{``#1''}%
\providecommand \bibnamefont  [1]{#1}%
\providecommand \bibfnamefont [1]{#1}%
\providecommand \citenamefont [1]{#1}%
\providecommand \href@noop [0]{\@secondoftwo}%
\providecommand \href [0]{\begingroup \@sanitize@url \@href}%
\providecommand \@href[1]{\@@startlink{#1}\@@href}%
\providecommand \@@href[1]{\endgroup#1\@@endlink}%
\providecommand \@sanitize@url [0]{\catcode `\\12\catcode `\$12\catcode
  `\&12\catcode `\#12\catcode `\^12\catcode `\_12\catcode `\%12\relax}%
\providecommand \@@startlink[1]{}%
\providecommand \@@endlink[0]{}%
\providecommand \url  [0]{\begingroup\@sanitize@url \@url }%
\providecommand \@url [1]{\endgroup\@href {#1}{\urlprefix }}%
\providecommand \urlprefix  [0]{URL }%
\providecommand \Eprint [0]{\href }%
\providecommand \doibase [0]{http://dx.doi.org/}%
\providecommand \selectlanguage [0]{\@gobble}%
\providecommand \bibinfo  [0]{\@secondoftwo}%
\providecommand \bibfield  [0]{\@secondoftwo}%
\providecommand \translation [1]{[#1]}%
\providecommand \BibitemOpen [0]{}%
\providecommand \bibitemStop [0]{}%
\providecommand \bibitemNoStop [0]{.\EOS\space}%
\providecommand \EOS [0]{\spacefactor3000\relax}%
\providecommand \BibitemShut  [1]{\csname bibitem#1\endcsname}%
\let\auto@bib@innerbib\@empty

\bibitem [{\citenamefont {Giuliani}\ and\ \citenamefont
  {Vignale}(2005)}]{Giuliani-book}%
  \BibitemOpen
  \bibfield  {author} {\bibinfo {author} {\bibfnamefont {G.}~\bibnamefont
  {Giuliani}}\ and\ \bibinfo {author} {\bibfnamefont {G.}~\bibnamefont
  {Vignale}},\ }\href@noop {} {\emph {\bibinfo {title} {Quantum theory of the
  electron liquid}}}\ (\bibinfo  {publisher} {Cambridge Univeristy Press},\
  \bibinfo {year} {2005})\BibitemShut {NoStop}%
\bibitem [{\citenamefont {Kazakov}\ \emph
  {et~al.}(2017{\natexlab{b}})\citenamefont {Kazakov}, \citenamefont {Simion},
  \citenamefont {Lyanda-Geller}, \citenamefont {Kolkovsky}, \citenamefont
  {Adamus}, \citenamefont {Karczewski}, \citenamefont {Wojtowicz},\ and\
  \citenamefont {Rokhinson}}]{Kazakov2017a-sup}%
  \BibitemOpen
  \bibfield  {author} {\bibinfo {author} {\bibfnamefont {Aleksandr}\
  \bibnamefont {Kazakov}}, \bibinfo {author} {\bibfnamefont {George}\
  \bibnamefont {Simion}}, \bibinfo {author} {\bibfnamefont {Yuli}\ \bibnamefont
  {Lyanda-Geller}}, \bibinfo {author} {\bibfnamefont {Valery}\ \bibnamefont
  {Kolkovsky}}, \bibinfo {author} {\bibfnamefont {Zbigniew}\ \bibnamefont
  {Adamus}}, \bibinfo {author} {\bibfnamefont {Grzegorz}\ \bibnamefont
  {Karczewski}}, \bibinfo {author} {\bibfnamefont {Tomasz}\ \bibnamefont
  {Wojtowicz}}, \ and\ \bibinfo {author} {\bibfnamefont {Leonid~P.}\
  \bibnamefont {Rokhinson}},\ }\bibfield  {title} {\enquote {\bibinfo {title}
  {Mesoscopic transport in electrostatically defined spin-full channels in
  quantum {H}all ferromagnets},}\ }\href {\doibase
  10.1103/PhysRevLett.119.046803} {\bibfield  {journal} {\bibinfo  {journal}
  {Phys. Rev. Lett.}\ }\textbf {\bibinfo {volume} {119}},\ \bibinfo {pages}
  {046803} (\bibinfo {year} {2017}{\natexlab{b}})}\BibitemShut {NoStop}%
\bibitem [{\citenamefont {Simion}\ \emph
  {et~al.}(2017{\natexlab{b}})\citenamefont {Simion}, \citenamefont {Kazakov},
  \citenamefont {Rokhinson}, \citenamefont {Wojtowicz},\ and\ \citenamefont
  {Lyanda-Geller}}]{Simion2017-sup}%
  \BibitemOpen
  \bibfield  {author} {\bibinfo {author} {\bibfnamefont {G.~E.}\ \bibnamefont
  {Simion}}, \bibinfo {author} {\bibfnamefont {A.}~\bibnamefont {Kazakov}},
  \bibinfo {author} {\bibfnamefont {L.~P.}\ \bibnamefont {Rokhinson}}, \bibinfo
  {author} {\bibfnamefont {T.}~\bibnamefont {Wojtowicz}}, \ and\ \bibinfo
  {author} {\bibfnamefont {Y.~B.}\ \bibnamefont {Lyanda-Geller}},\ }\bibfield
  {title} {\enquote {\bibinfo {title} {Disorder-generated non-{A}belions},}\
  }\href@noop {} {\  (\bibinfo {year} {2017}{\natexlab{b}})},\ \Eprint
  {http://arxiv.org/abs/arXiv:1707.02929} {arXiv:1707.02929} \BibitemShut
  {NoStop}%
\bibitem [{\citenamefont {{MacDonald}}(1990)}]{macdonald_edge_1990}%
  \BibitemOpen
  \bibfield  {author} {\bibinfo {author} {\bibfnamefont {A.~H.}\ \bibnamefont
  {{MacDonald}}},\ }\bibfield  {title} {\enquote {\bibinfo {title} {Edge states
  in the fractional quantum {H}all effect regime},}\ }\href@noop {} {\bibfield
  {journal} {\bibinfo  {journal} {Physical review letters}\ }\textbf {\bibinfo
  {volume} {64}},\ \bibinfo {pages} {220} (\bibinfo {year} {1990})}\BibitemShut
  {NoStop}%
\bibitem [{\citenamefont {Beenakker}(1990)}]{beenakker_edge_1990}%
  \BibitemOpen
  \bibfield  {author} {\bibinfo {author} {\bibfnamefont {C.~W.~J.}\
  \bibnamefont {Beenakker}},\ }\bibfield  {title} {\enquote {\bibinfo {title}
  {Edge channels for the fractional quantum {H}all effect},}\ }\href@noop {}
  {\bibfield  {journal} {\bibinfo  {journal} {Physical review letters}\
  }\textbf {\bibinfo {volume} {64}},\ \bibinfo {pages} {216} (\bibinfo {year}
  {1990})}\BibitemShut {NoStop}%
\bibitem [{\citenamefont {Wen}\ and\ \citenamefont
  {Zee}(1991)}]{wen_topological_1991}%
  \BibitemOpen
  \bibfield  {author} {\bibinfo {author} {\bibfnamefont {X.~G.}\ \bibnamefont
  {Wen}}\ and\ \bibinfo {author} {\bibfnamefont {A.}~\bibnamefont {Zee}},\
  }\bibfield  {title} {\enquote {\bibinfo {title} {Topological structures,
  universality classes, and statistics screening in the anyon superfluid},}\
  }\href@noop {} {\bibfield  {journal} {\bibinfo  {journal} {Physical Review
  B}\ }\textbf {\bibinfo {volume} {44}},\ \bibinfo {pages} {274} (\bibinfo
  {year} {1991})}\BibitemShut {NoStop}%
\bibitem [{\citenamefont {Wen}(1992)}]{wen_theory_1992}%
  \BibitemOpen
  \bibfield  {author} {\bibinfo {author} {\bibfnamefont {Xiao-Gang}\
  \bibnamefont {Wen}},\ }\bibfield  {title} {\enquote {\bibinfo {title} {Theory
  of the edge states in fractional quantum {H}all effects},}\ }\href@noop {}
  {\bibfield  {journal} {\bibinfo  {journal} {International journal of modern
  physics B}\ }\textbf {\bibinfo {volume} {6}},\ \bibinfo {pages} {1711--1762}
  (\bibinfo {year} {1992})}\BibitemShut {NoStop}%
\bibitem [{\citenamefont {Chamon}\ and\ \citenamefont
  {Wen}(1994)}]{chamon_sharp_1994}%
  \BibitemOpen
  \bibfield  {author} {\bibinfo {author} {\bibfnamefont {C.~de~C.}\
  \bibnamefont {Chamon}}\ and\ \bibinfo {author} {\bibfnamefont {X.~G.}\
  \bibnamefont {Wen}},\ }\bibfield  {title} {\enquote {\bibinfo {title} {Sharp
  and smooth boundaries of quantum {H}all liquids},}\ }\href@noop {} {\bibfield
   {journal} {\bibinfo  {journal} {Physical Review B}\ }\textbf {\bibinfo
  {volume} {49}},\ \bibinfo {pages} {8227} (\bibinfo {year}
  {1994})}\BibitemShut {NoStop}%
\bibitem [{\citenamefont {Wu}\ \emph {et~al.}(2012)\citenamefont {Wu},
  \citenamefont {Sreejith},\ and\ \citenamefont {Jain}}]{Jainedges}%
  \BibitemOpen
  \bibfield  {author} {\bibinfo {author} {\bibfnamefont {Ying-Hai}\
  \bibnamefont {Wu}}, \bibinfo {author} {\bibfnamefont {G.~J.}\ \bibnamefont
  {Sreejith}}, \ and\ \bibinfo {author} {\bibfnamefont {Jainendra~K.}\
  \bibnamefont {Jain}},\ }\bibfield  {title} {\enquote {\bibinfo {title}
  {Microscopic study of edge excitations of spin-polarized and spin-unpolarized
  $\nu=2/3$ fractional quantum {H}all effect},}\ }\href@noop {} {\bibfield
  {journal} {\bibinfo  {journal} {Phys. Rev. B}\ }\textbf {\bibinfo {volume}
  {86}},\ \bibinfo {pages} {115127} (\bibinfo {year} {2012})}\BibitemShut
  {NoStop}%
\bibitem [{\citenamefont {Chklovskii}(1995)}]{PhysRevB.51.9895}%
  \BibitemOpen
  \bibfield  {author} {\bibinfo {author} {\bibfnamefont {Dmitri~B.}\
  \bibnamefont {Chklovskii}},\ }\bibfield  {title} {\enquote {\bibinfo {title}
  {Structure of fractional edge states: A composite-fermion approach},}\
  }\href@noop {} {\bibfield  {journal} {\bibinfo  {journal} {Phys. Rev. B}\
  }\textbf {\bibinfo {volume} {51}},\ \bibinfo {pages} {9895--9902} (\bibinfo
  {year} {1995})}\BibitemShut {NoStop}%
\bibitem [{\citenamefont {McDonald}\ and\ \citenamefont
  {Haldane}(1996)}]{mcdonald_topological_1996}%
  \BibitemOpen
  \bibfield  {author} {\bibinfo {author} {\bibfnamefont {I.~A.}\ \bibnamefont
  {McDonald}}\ and\ \bibinfo {author} {\bibfnamefont {F.~D.~M.}\ \bibnamefont
  {Haldane}},\ }\bibfield  {title} {\enquote {\bibinfo {title} {Topological
  phase transition in the $\nu½= 2/3$ quantum {H}all effect},}\ }\href@noop {}
  {\bibfield  {journal} {\bibinfo  {journal} {Phys. Rev. B}\ }\textbf {\bibinfo
  {volume} {53}},\ \bibinfo {pages} {15845} (\bibinfo {year}
  {1996})}\BibitemShut {NoStop}%
\bibitem [{\citenamefont {Moore}\ and\ \citenamefont
  {Haldane}(1997)}]{moore_edge_1997}%
  \BibitemOpen
  \bibfield  {author} {\bibinfo {author} {\bibfnamefont {J.~E.}\ \bibnamefont
  {Moore}}\ and\ \bibinfo {author} {\bibfnamefont {F~D~M}\ \bibnamefont
  {Haldane}},\ }\bibfield  {title} {\enquote {\bibinfo {title} {Edge
  excitations of the $\nu=2/3$ spin-singlet quantum {H}all state},}\
  }\href@noop {} {\ \textbf {\bibinfo {volume} {55}},\ \bibinfo {pages} {7818}
  (\bibinfo {year} {1997})}\BibitemShut {NoStop}%
\bibitem [{\citenamefont {Haldane}(1983)}]{haldane83-sup}%
  \BibitemOpen
  \bibfield  {author} {\bibinfo {author} {\bibfnamefont {F.D.M.}\ \bibnamefont
  {Haldane}},\ }\bibfield  {title} {\enquote {\bibinfo {title} {Fractional
  quantization of the {H}all effect: a hierarchy of incompressible quantum
  fluid states},}\ }\href@noop {} {\bibfield  {journal} {\bibinfo  {journal}
  {Physical Review Letters}\ }\textbf {\bibinfo {volume} {51}},\ \bibinfo
  {pages} {605 -- 8} (\bibinfo {year} {1983})}\BibitemShut {NoStop}%
\bibitem [{\citenamefont {Clarke}\ \emph
  {et~al.}(2012{\natexlab{b}})\citenamefont {Clarke}, \citenamefont {Alicea},\
  and\ \citenamefont {Shtengel}}]{Clarke2012-sup}%
  \BibitemOpen
  \bibfield  {author} {\bibinfo {author} {\bibfnamefont {David~J.}\
  \bibnamefont {Clarke}}, \bibinfo {author} {\bibfnamefont {Jason}\
  \bibnamefont {Alicea}}, \ and\ \bibinfo {author} {\bibfnamefont {Kirill}\
  \bibnamefont {Shtengel}},\ }\bibfield  {title} {\enquote {\bibinfo {title}
  {Exotic non-{A}belian anyons from conventional fractional quantum {H}all
  states},}\ }\href {\doibase doi:10.1038/ncomms2340} {\bibfield  {journal}
  {\bibinfo  {journal} {Nat. Commun.}\ }\textbf {\bibinfo {volume} {4}},\
  \bibinfo {pages} {1348} (\bibinfo {year} {2012}{\natexlab{b}})}\BibitemShut
  {NoStop}%
\end{thebibliography}
\end{document}